\documentclass[aps,physrev,graphicx,amsmath,amssymb,reprint,superscriptaddress]{revtex4-2} 
\usepackage{graphicx}
\usepackage{mathtools}
\usepackage{orcidlink}

\usepackage[version=3]{mhchem} 

\usepackage{dcolumn}
\newcolumntype{d}[1]{D{.}{\cdot}{#1} }

\usepackage{bm}
\usepackage{xcolor,soul}

\usepackage{siunitx}

\begin{document}
\title{Correcting hybrid density functionals to model Y6 and other non-fullerene acceptors}
\date{\today} 

\author{Tom Ward *\,\orcidlink{0009-0009-0658-0740}}
\thanks{These two authors contributed equally}
\affiliation{Department of Chemistry, Imperial College London, Exhibition Road, London  SW7 2AZ, UK}

\author{Isabel Creed ,\orcidlink{0009-0001-3930-7873}}
\thanks{These two authors contributed equally}
\affiliation{Department of Chemistry, Imperial College London, Exhibition Road, London  SW7 2AZ, UK}
\thanks{These two authors contributed equally}

\author{Tim Rein \orcidlink{0009-0004-9681-0055}}
\affiliation{Department of Physics, Imperial College London, Exhibition Road, London  SW7 2AZ, UK}
\affiliation{Department of Chemistry, Imperial College London, Exhibition Road, London  SW7 2AZ, UK}

\author{Jarvist Moore Frost\,\orcidlink{0000-0003-1938-4430}}
\affiliation{Department of Chemistry, Imperial College London, Exhibition Road, London  SW7 2AZ, UK}
\affiliation{Department of Physics, Imperial College London, Exhibition Road, London  SW7 2AZ, UK}

\email[Electronic mail:]{jarvist.frost@imperial.ac.uk}


\begin{abstract}
Recently developed fused-ring electron-acceptors such as Y6 (BTP-4F) have strong oscillator strength, good charge-carrier transport and a small bandgap. They therefore have enormous current technical application to organic optoelectronics, such as solar cells. To design new materials, it would be useful to predict the electronic structure accurately. Due to the large number of atoms involved in representative aggregates of these materials, we need an efficient electronic structure method. Standard density functional theory poorly describes charge-transfer states, and were typically parameterised for vacuum calculations of individual molecules. 

In this work we tune a range-separated hybrid functional for Y6, and characterise representative dimers extracted from the solid-state. We demonstrate that the extensive solvatochromic effects of Y6 are due, in part, to oscillator strength borrowing between the charge-transfer and Frenkel excitons. We provide an explanation for the short optimally-tuned range-separation parameter, based on the Penn model for the frequency dependent dielectric of a semiconductor. We caution that non-tuned range-separated hybrids are less accurate than global hybrids for these, and similar, materials. We show how reducing the range-separation length improves the accuracy of standard range-separation functionals, without an involved tuning process. 

   
\end{abstract}

\maketitle 

\section{Introduction}\label{introduction}
The development of non-fullerene (electron) acceptors (NFAs) has enabled
a significant step forwards in organic photovoltaic (OPV) power-conversion
efficiency. 
Modern fused-ring electron-acceptors (FREAs) have strong absorption, tuned to the near-infrared part of the solar spectrum using donor-acceptor (push-pull) molecular design. 

This strong oscillator strength and small band-gap, packed into a small volume,
directly leads to a large dielectric constant, which has key ramifications for
the correct quantum-chemical methods. 

The archetypical modern FREA is Y6 (BTP-4F)\cite{Yang2021}. 
Y6 has a banana-shaped, but flat, A-DA'D-A structure, with the alkyl side-chains,
required for solubility, kept apart from the solid-state $\pi-\pi$ stacking of
the fused cores required for high inter-molecular mobility. 
The central unit is electron-deficient benzothiadiazole (A') fused with
electron-rich thieno[3,2-b]thiophene-pyrole (D), linked via conjugated vinyl
bonds to electron-poor fluorinated and cyano-substituted indanone (A)\cite{yuan2019single}. 

Y6 exhibits good processing behaviour with a wide variety of co-deposited
materials, has strong optical absorption of its own, and seems to possess
good charge-carrier transport properties\cite{yuan2019single}. 
Overall, Y6 has become the first choice electron-acceptor in a bulk heterojunction organic photovoltaic. 

Neat Y6 films demonstrate unusual photophysics. 
The solvatochromic shift of Y6 in the solid-state is giant. 
A computational study\cite{Han2020} predicts that this giant shift is only seen in the singlet state energies,
whereas the modest shift of the triplet state energies are consistent with other FREAs. 
There is some evidence that free charges may be directly generated without an
applied bias\cite{price2022free}; certainly the energetic cost of exciton
dissociation into free charges is small\cite{Zhu2022,Hart2026}. 
The separation of tightly-bound Frenkel-like excitons in organic semiconductors into
free charges is the central theoretical mystery of the operation of organic
photovoltaics: a na\"ive model considering a bulk dielectric constant
would suggest an exciton binding energy of over an electron Volt. 

As a simple bilayer with rubrene, Y6 upconverts infrared light into visible (via triplet-triplet annihilation; two photons in for one photon out) with the highest efficiency for a solid-state device\cite{izawa2021efficient}.

Often the concept of intermolecular charge transfer (CT)
states\cite{si2023photo, price2022free,izawa2021efficient, lan2024correction}
is invoked to explain the unusual photophysics of Y6. 
As these are intermolecular properties, 
a theoretical treatment first requires a structural model. 
Initially this required molecular dynamic simulations\cite{Zhang2020},
but more recent studies\cite{giannini2024role} often use the now available high quality crystal structures\cite{Xiao2020} (CSD code OHUBUR).

Charge-transfer states are challenging to describe well with density functional theory (DFT). 
The self-interaction error in DFT is often corrected by including a fraction of exact exchange from Hartree-Fock. 
But as these `global' hybrids have the incorrect long-range behaviour, CT states can become qualitatively incorrect. 

Optimally tuned screened range-separated hybrid (OT-SRSH) functionals were developed explicitly to have a well balanced description of both local and long-range excitations\cite{RefaelyAbramson2013,kummel2017charge, bhandari2019quantitative, zheng2017effect}. 
They achieve this by having the fraction of Hartree-Fock exchange follow an
error function, transitioning from the locally screened ($\alpha$
fraction of Hartree-Fock) domain to a long-range (screened by only the macroscopic
dielectric constant, $\alpha+\beta$ Hartree-Fock fraction). 
For vacuum range-separated hybrid (RSH) functionals (assuming $\epsilon_\infty=1$ and
constrained by $\alpha+\beta=1$) this recovers the full Hartree-Fock limit at long
range; whereas for \textit{screened} range-separated hybrid (SRSH) functionals
($\epsilon_\infty>1$), the long-range fraction is reduced by the macroscopic dielectric screening. 

Here we tune the LC-$\omega$hPBE functional. 
This was previously indicated as having high accuracy for Y5 and ITIC monomers by Franco et al.\cite{Franco2023}; used by Giannini et al.\cite{giannini2024role} to model Y6 dimers; and more recently applied to excitons in solid-state Y6 by Akram et al.\cite{akram2025analyzing}.
This functional has been used to model organic heterojunctions such as the C$_{60}$/Pentacene and P3HT/PCBM dimer\cite{zheng2017effect}. This functional is believed to predict the correct ordering of the CT and FE states in molecular systems\cite{kummel2017charge, bhandari2019quantitative, zheng2017effect}. 

We analyse the excited states predicted by our tuned LC-$\omega$hPBE
functional, and compare to untuned range-separated and global-hybrid functionals. 
We find that the optimal fits produce a \textit{negative} beta parameter, leading to a functional where Hartree-Fock is \textit{more} screened at longer range. 
This we link back via a GW picture to the high charge-carrier mobility of the material. 
Using the simple Penn\cite{Penn1962} model for the dielectric function of a semiconductor, we can predict the range separation parameter directly from the small bandgap. 

Our key warning is that `off the shelf' range-separated hybrid functionals (such as CAM-B3LYP), which have been developed for typical organic molecules, are actually worse at predicting excited states in modern low-bandgap high-absorbance organic semiconductors than the global hybrids (such as B3LYP). 
We find that by modifying the range-separation parameter of the popular CAM-B3LYP functional, to reflect the small band gap and large dielectric constant of Y6, we are able to reproduce the same correct excited state properties, ranking, and absolute energy as predicted by our much more involved optimal-tuning of LC-$\omega$hPBE. 
Our key recommendation is that when modelling low-bandgap large-oscillator-strength high-dielectric materials (i.e. all modern organic semiconductors), that a suitably reduced range-separation parameter is used. 

\section{Methods and Computational Details}\label{OTSRSH Methods}

\subsection{Optimally tuned range-separated hybrid}

Optimally-tuned (OT) range-separated hybrid functionals empirically impose an Ewald partition of the  Coulomb energy operator\cite{zheng2017effect,bhandari2018fundamental},

\begin{equation}
    \frac{1}{r} = \frac{\alpha+\beta\cdot \text{erf}(\omega r)}{r} +\frac{1-(\alpha+\beta\cdot \text{erf}(\omega r))}{r}.
    \label{OT-SRSH}
\end{equation}

The $\alpha$,  $\beta$, and $\omega$ parameters are then adjustable for the system of interest\cite{zhou2021range}. 
The parameter $\omega$ determines the distance at which one transfers from
short-range exchange fraction ($\alpha$), to long-range exchange fraction
($\alpha+\beta$). 

For a given value of $\omega$, the exchange correlation energy is an admixture as\cite{zhou2021range}

\begin{equation} \begin{aligned} E_{XC}^{SRSH} = & \ E_{PBE(C)} \\ & + \alpha E_{HF(X)}^{SR} + (1-\alpha)E_{PBE(X)}^{SR} \\ & + (\alpha+\beta)E_{HF(X)}^{LR}  +(1-(\alpha+\beta))E_{PBE(X)}^{LR}. \end{aligned} 
\label{EXC OT-SRSH}
\end{equation}

The terms in the summation of Eq.\ \ref{EXC OT-SRSH} are: (global) GGA/PBE
correlation energy; short range (`SR') HF and GGA/PBE exchange; long range
(`LR') HF exchange and GGA/PBE exchange.

Tuning is achieved by enforcing a generalisation of Koopmans' theorem, where the energy of the Kohn-Sham frontier orbitals reproduces the ionisation potential (IP) and electron affinity (EA) of the system, as calculated consistently with a delta-SCF method and the same functional. 
As proposed by Zheng et al.\cite{zheng2017effect}, we optimise HOMO and LUMO eigenvalues simultaneously, minimising a mean squared loss 
\begin{equation}
    J(\omega) = [\epsilon_{HOMO}(\omega) + IP(\omega) ]^2 + [\epsilon_{LUMO}(\omega) + EA(\omega)]^2.
    \label{omega function}
\end{equation}

We use LC-$\omega$hPBE, a screened range-separated hybrid (SRSH) functional, and so constrain $\alpha+\beta\equiv1/\epsilon_r$\cite{zheng2017effect,bhandari2018fundamental} when minimising $J(\alpha,\beta)$, where $\epsilon_r$ is the bulk optical dielectric constant of the system, having first fitted $\omega$ in a vacuum. 
This allows us to impose the macroscopic behaviour of the solid-state material, without requiring this to be perfectly recovered from the functional, and not requiring a noisy fitting of $\omega$ in a polarisable medium. 
This is in contrast to tuning for RSH functionals, where the constraint $\alpha+\beta\equiv1$ is applied after having fitted $\omega$ in a polarisable medium. 

Measured values of $\epsilon_r$ for Y6 films range from $\epsilon_r\simeq 2.5$\cite{epsilon25} to $\epsilon_r \simeq 6$\cite{epsilon573} to $\epsilon_r\simeq 11$\cite{epsilon108}. 
Following discussion with experimental colleagues, we chose what we felt was a high but still well-founded value of $\epsilon_r \simeq 6$. 
We consider the effect of modifying $\epsilon_r=3-10$ in the supplementary information (figure \ref{epsilon}) and found that the ordering and other properties of the low lying CT and FE singlet and triplet states do not depend strongly on the value of the dielectric constant chosen in our tuning.

\subsection{Wavefunction Analysis}
To demonstrate that the OT-SRSH functional accurately describes excited states in Y6, we used TheoDORE\cite{plasser2020theodore} to analyse the character of each active state. 
TheoDORE calculates state properties by considering the ratio of the diagonal and off-diagonal elements of, $D_{ab}^{0i}$, the ground to $i$-th excited state transition density matrix (TDM) in the orbital basis $\{a,b\}$. The orbitals $\{a,b\}$ are localised on the molecular fragments $A$ and $B$ \cite{plasser2020theodore}. 

For the Y6 dimers, a natural motif for the fragments is the two monomers. One can further naturally subdivide Y6 into two acceptor end units (A) and a donor core (A'DA').

Two key properties we want to predict are: the charge-transfer (CT) character; and the extent of localisation. 

In TheoDORE\cite{plasser2012analysis,plasser2020theodore}, the CT character is given by the parameter $\omega_{CT}$ which corresponds to the proportion of the excitation which occurs between fragments---the degree to which the excitation leads to charge transfer. 
A value of $\omega_{CT}=1$ therefore corresponds to a `pure' CT state, while $\omega_{CT} = 0$ is a `pure` Frenkel-Exciton (CT) state. 
Due to CT-FE state mixing at short separations most excited states of the dimers are not pure FE or CT states. 
For Y6, this has a considerable effect on the energetic ordering of the singlet and triplet states. 
Following Zhang et al.\cite{zheng2017effect}, we consider CT states as those states with $\omega_{CT} \gtrsim 0.75$ and FE states as those states with $\omega_{CT} \lesssim 0.25$ with all other states being considered ``mixed'' states . 

We define the excitation localisation by the participation ratio, $\omega_{PR}$ \cite{plasser2020theodore}. 
For two Y6 monomers, $\omega_{PR}= 2$ corresponds to an excited state wavefunction that is entirely delocalised (evenly distributed) across both monomers, while; $\omega_{PR}=1 $
describes full localisation on one monomer.

\subsection{Computational details}

Gaussian16\cite{g16} was used for all electronic structure calculations in this work. 
We use a modest 6-31G(d,p) basis set. 
This basis set was chosen after checking basis-set dependence for excited state energies (shown in the supplementary information in figure \ref{converge}), and tests indicating that calculations converged reliably and smoothly. 

We studied the Y6 monomer and a set of six contact pair dimers; 
these dimers were extracted from a solved crystal structure\cite{xiao2020single}, with a distance cut-off. 
Among the unique pairs, eight have a minimum separation of less than 5 \si{\angstrom}, defined by Giannini et al.\cite{giannini2024role} as `contact pair dimers', of which we study D1-D6 which have a minimum intermolecular distance of less than 3.5 Å. 
At these small separations, intermolecular electron transfer integrals become significant and affect the excited state energies\cite{olaya2011energy}.

\section{Results and Discussion}

\subsection{OT-SRSH for Y6 dimers}

The tuned parameters for representative Y6 dimers, and the monomer, alongside the defaults, are given
in Table \ref{tuning}. 
The parameters for different dimers are within 10\% of each other, and the monomer tuned parameters are similarly close. 
This suggests that the tuned parameters are robust to different dimer configuration. 
As we only tune on contact dimers in a PCM, we cannot make a strong statement about whether these parameters would also be correct for long-range pairs. 

A key observation is that the $\beta$ parameter is negative: screening is \textit{greater} at longer range. 
Though unusual for molecular systems, this is entirely consistent with the formalism, and has been observed before\cite{Chai2008}. 
Here we connect this to the point of view that the high charge carrier mobility in Y6 provides additional macroscopic screening, like a traditional semiconductor. 

\begin{table}[h!]
\centering
\begin{tabular}{|l|c|c|c|c|}
\hline
\textbf{Dimer} & Type & \textbf{$\omega$} ($a_0^{-1}$) & \textbf{$\alpha$}
    & \textbf{$\beta$} \\
\hline
Default & N/A      & 0.40  & 0      & 1 \\
Monomer & -        & 0.140 & 0.260 & -0.093 \\
D1 & Contact       & 0.120 & 0.230 & -0.063 \\
D2 & Contact       & 0.110 & 0.250 & -0.083 \\
D3 & Contact       & 0.110 & 0.240 & -0.073 \\
D4 & Contact       & 0.120 & 0.250 & -0.083 \\
D5 & Contact       & 0.120 & 0.240 & -0.073 \\
D6 & Contact       & 0.120 & 0.240 & -0.073 \\
D7 & Contact       & 0.130 & 0.250 & -0.083 \\
D8 & Contact       & 0.130 & 0.250 & -0.083 \\
D9 & Non-Contact   & 0.130 & 0.250 & -0.083 \\
D10 & Non-Contact  & 0.130 & 0.250 & -0.083 \\
\hline
\end{tabular}
\caption{LC-$\omega$hPBE default parameters, and then our tuned LC-$\omega$hPBE functional parameters for the Y6 monomer and a sample of Y6 dimers \cite{giannini2024role}.  
The units of $\omega$ are inverse Bohr radius, $a_0^{-1}$, whereas $\alpha$ and $\beta$ are unitless fractions. 
The 6-31G(d,p) basis set was used throughout. 
Tuning of $\alpha$ and $\beta$ was completed in a PCM (using the SCRF method) with $\epsilon_r = 6$. 
Note that our predicted $\beta$ fraction is slightly \textit{negative}, this means that there is less Hartree-Fock exchange at long range than at short range. 
The ensemble average of $\omega$ for the dimers is 0.122, consistent with a recently published value for Y6 aggregates of 0.127\cite{akram2025analyzing}.}
\label{tuning}
\end{table}

We benchmark the excited state energies from our tuned functional, and standard methods, against GW/BSE/MM calculations\cite{giannini2024role}.

We present the mean-squared error for the first six singlet excited states in Table \ref{GW_compare}. 
Error is minimised for our tuned SRSH (LC-$\omega$hPBE); it is very notable that non-tuned range-separated hybrids demonstrate the largest error, significantly larger than the global hybrids. 

\begin{table}
\begin{center}
\begin{tabular}{||c |c|c|c|c|c|c|} 
 \hline
 Functional  & D1 & D2& D3 &D4 & D5 & D6  \\ 
 \hline
 LC-$\omega$hPBE & 0.04 & 0.06 &0.05& 0.09&0.09&0.12\\
 \hline
 B3LYP & 0.05 &0.10&0.09&0.11&0.09&0.12 \\
 \hline
 PBE0 & 0.10&0.08&0.08&0.11&0.14&0.16 \\
 \hline
 CAM-B3LYP & 0.64 &0.56&0.56&0.69&0.72&0.75 \\ 
 \hline
 wB97XD & 0.76  & 0.67&0.67&0.82&0.87&0.92\\
\hline 
\end{tabular}
\label{table_energydiff}
\end{center}
\caption{Table of mean-squared error (versus GW/BSE/MM references) $\Delta_S^{TDDFT}$ (eq. \ref{diff}) for the first six excited state singlets, in eV.  }
\label{GW_compare}
\end{table}

In the supplementary information (figs.\  \ref{D1_state}, \ref{D2_state} and
\ref{D4_state}), we show how the energy (E) and  extent of charge transfer
($\omega_{CT}$) of the six lowest singlet and triplet states of the D1, D2 and
D4 dimers depend on the functional. 

The functionals can be classified based on the ordering of the FE, CT, and mixed CT-FE states. 
The OT-SRSH, B3LYP and PBE0 functionals produce a state ordering consistent with the GW/BSE results (private communication with authors of \cite{giannini2024role}: they observe no state crossing of the dimer excited states from the PBE0 starting point to the eigenvalue self-consistent GW/BSE/MM results). 
The default parameter range-separated functionals CAM-B3LYP and wB97XD give an erroneous ordering.

B3LYP is known to underestimate the energy of CT states, particularly on dissociation of a dimer\cite{magyar2007dependence}.
The OT-SRSH functional is expected\cite{kummel2017charge, bhandari2019quantitative, zheng2017effect} to correctly order CT and FE states. 
Therefore, agreement of our OT-SRSH to B3LYP initially gave us concern.  
In the supplementary information (figs.\ \ref{dimer_seperation_OT} and  \ref{dimer_seperation_B3LYP}) we show that the OT-SRSH functional is able to obtain the correct behaviour of the ordering of the CT and FE states during dimer dissociation, whereas B3LYP does not. It is a happy coincident that in these dimers the global hybrids are correctly balanced.

For our tuned SRSH to be used in the study of Y6 solid and thin films, the tuning parameters must be transferable, i.e. the same parameters should work consistently well on monomers, dimers, and larger aggregates. 
This is to avoid the need for repeating the tuning process on different structures, and to avoid having to extrapolate from increasingly large aggregates to the thermodynamic limit. 
To demonstrate transferability of the tuned functional, we compare the singlet and triplet excited state properties for all contact pair dimers when: the calculations are run using the tuned parameters for dimer being considered; and when the calculations are run using the monomer tuning. 
We find that the state properties in both cases are practically identical. 
In the supplementary information (fig.\ \ref{D2_size_extensivity}) we show detailed analysis of this for D2 dimer, with similar results being found for all dimers examined. 

A small reorganisation energy enables large charge mobility\cite{Wang2011, Lin2019}, and reduces non-radiative losses in OPVs \cite{Azzouzi2018}. 
Given the importance of the reorganisation energy we validate the use of our
OT-SRSH in calculating the reorganisation energy of the monomer using both Nelsen's four-point method \cite{Nelsen1987} and Reimers' normal-mode-resolved approach \cite{Reimers2001}.  
These data we show in figures \ref{fig:4P_RE_Excited_Charged_States} and \ref{fig:dushin_spectra_normalised} of the supplementary information. 
We find that our tuned SRSH has similar values of reorganisation energy to the commonly used functionals PBE0 and B3LYP; our tuning for excited states has not made the prediction of the potential energy surfaces erroneous. 
All tested functionals are capable of capturing the relatively small reorganisation energy of the Y6 first-singlet, which is notably smaller than for the cation or anion.

\subsection{Excited state properties of Y6}

\begin{figure}
    \centering
    \includegraphics[width=\linewidth]{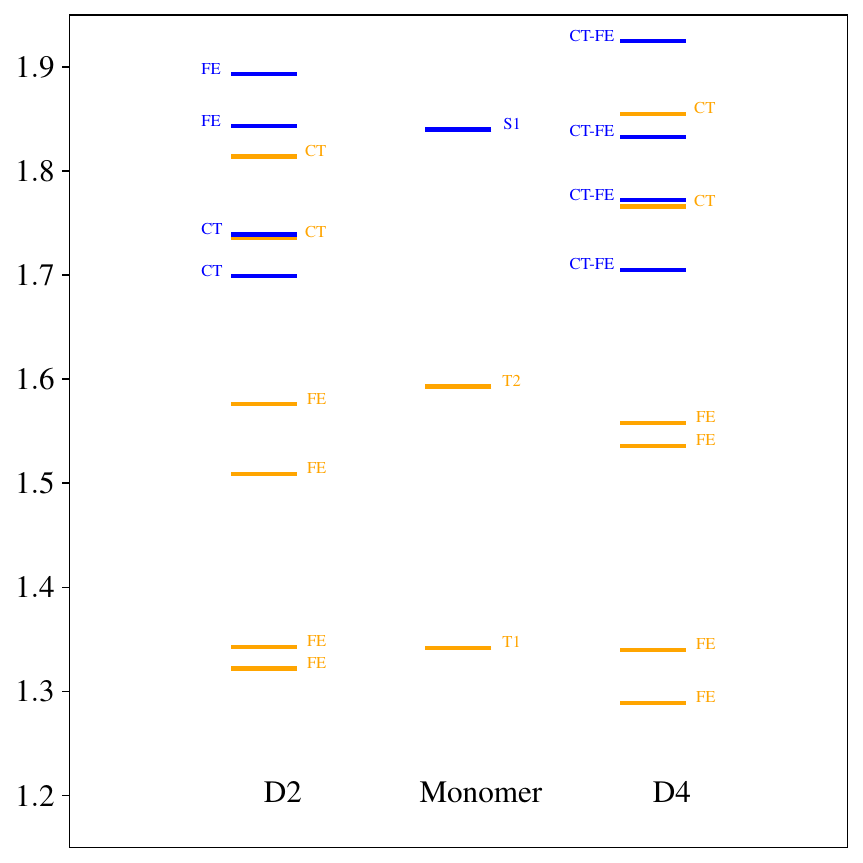}
    \caption{Jablonski diagram of the D2 dimer (example of J aggregate), D4 dimer (example of H aggregate) and monomer. Triplet states are in orange and the Singlet states are in blue. }
    \label{Jablonski}
\end{figure}

Having somewhat validated our OT-SRSH functional, we now move on to considering the properties of the excited states of the contact dimers. 
Our full analysis of the excited-state properties of the dimers is in the supplementary information (Figs. \ref{excited_properties}, \ref{excited_properties_2} and \ref{e/h correlation}). 

Dimers D2 and D4 represent two distinct dimer behaviours in Y6 crystal. 
We show their Jablonski diagram alongside the monomer in Fig.~\ref{Jablonski}. 

In the D2 dimer the two lowest-in-energy singlet states are CT states followed by two higher-in-energy FE states, while for the D4 dimer all four lowest energy singlet states are mixed CT-FE states. 
The strong dependence of the FE-CT mixing in the singlet states on the class of dimer is shown explicitly by examining $\omega_{CT}$ in the supplementary information (Figs. \ref{excited_properties} and \ref{excited_properties_2}). 
By contrast, the four lowest-in-energy triplet states are FE states, followed by two higher-in-energy CT states.  
in the supplementary information we show (in figures \ref{excited_properties} and \ref{excited_properties_2}) that the triplet state FE-CT mixing has a much smaller dependence on the type of dimer than the singlet states.  
We find that the extent of delocalisation of the triplet excited states, as probed by $\omega_{PR}$, is $\omega_{PR}\simeq 2$ in  the D2 dimer to $\omega_{PR} \simeq 1$ in the D4 dimer, whereas singlet states are consistently delocalised ($\omega_{PR} \sim 2$). 

The triplet CT states of both dimers lie higher in energy than the singlet CT state/mixed CT-FE states. This is because the singlet CT states, are stabilised by this mixing. 
In contrast, triplet CT states are de-stabilised to higher energy states. 
Similar behaviour is observed in interfacial dimers of Y6:ZR1\cite{miao2024energetic}, where it was proposed that significant CT-FE mixing causes singlet-triplet inversion. 
This inversion of singlet/triplet interfacial CT states has been hypothesised to reduce recombination in Y6 OPVs. 

The key role of the CT-FE mixing in the bathochromatic shift of the singlet
states has been noted\cite{giannini2024role}. However, in this study it is predicted that the
CT states of the dimer lie above the FE states. 
This means that they failed to predict the existence of the polaronic states of charge located on neighbouring Y6 molecules, believed to be present in their experimental data \cite{giannini2024role}. 

Our tuned OT-SRSH predicts that these spectroscopic features are a charge-transfer state
formed by contact dimers.  
There is some evidence for such low-lying CT states being observed in electroabsorption (EA) measurements\cite{mahadevan2024assessing}. 

\subsection{Fixing range-separated hybrids without tuning}

\begin{figure*}
    \centering
    \includegraphics[width=1\linewidth]{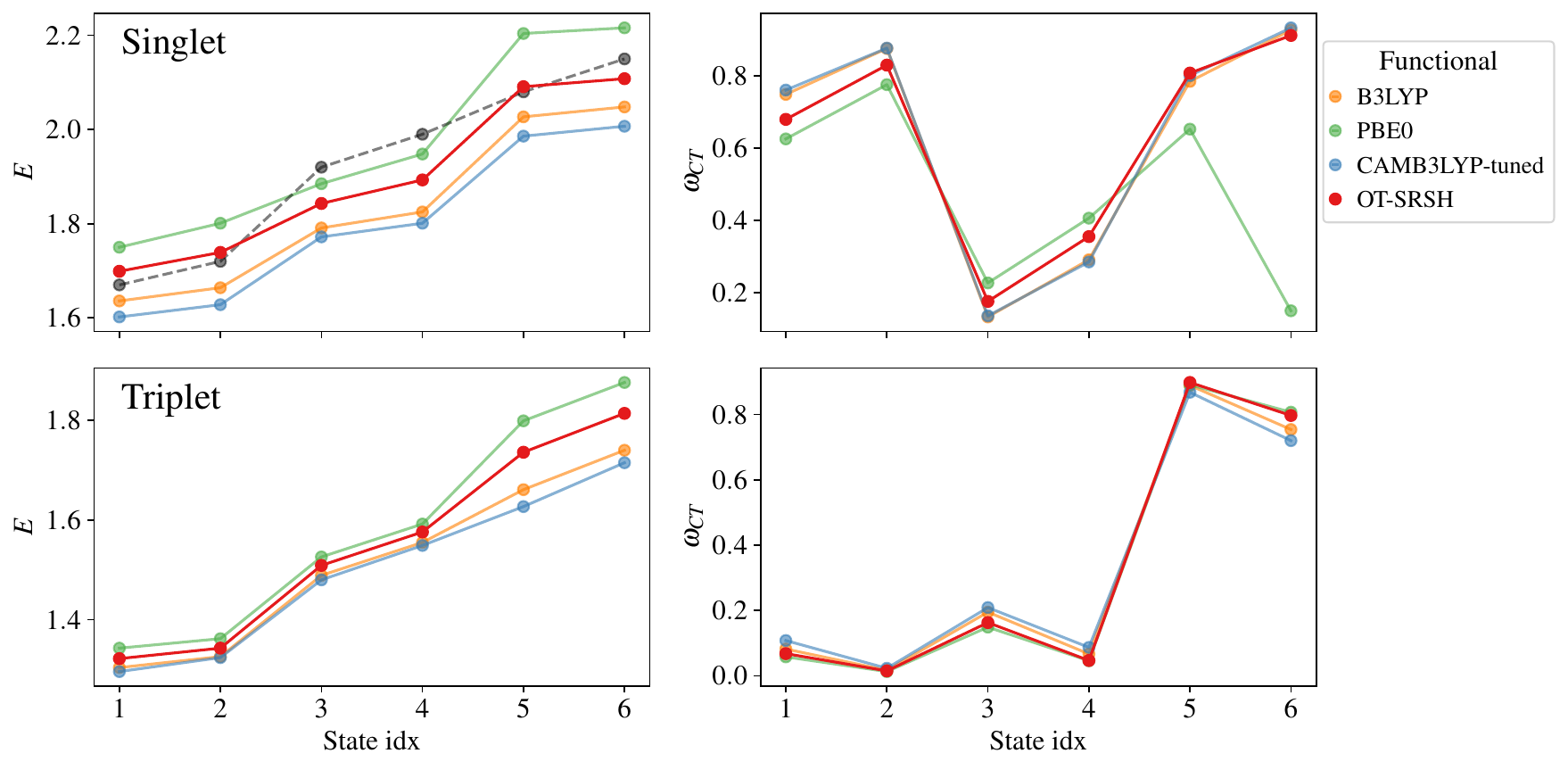}
    \caption{Comparison of the energies ($E$) and extent of charge transfer ($\omega_{CT}$) given by OT-SRSH, B3LYP, PBE0 and CAM-B3LYP-tuned (CAM-B3LYP with a changed range-separated parameter).}
    \label{tuned_dielectric}
\end{figure*}

Our tuned parameters are consistent between the monomer and dimers. The key change is the larger dielectric constant screening the Coulomb interaction (thus reducing Hartree-Fock exchange), and a shortened range-separation parameter.

GW theory calculates the full dielectric function (frequency dependent) in the Random Phase Approximation (RPA) and then uses this to screen the Coulomb (Hartree-Fock) interaction. 

Hybrid density functionals can be considered a static approximation to this,
where the fraction of exact exchange (an empirical parameter) then presents as being $1/\epsilon_\infty$. 
From this perspective, PBE0 ($25\%$ exact exchange) can be interpreted as
implicitly assuming an optical dielectric constant of 4. 
The interpretation for functionals with more complex admixtures of exchange-correlation (such as B3LYP) is not so clear.

Screened range-separated hybrid density functionals can be considered an extension possessing a model of the inverse dielectric function: that of the error function. 
At very short distances (high frequencies) the dielectric response of the material is not fast enough to screen the Coulomb interaction; at large distances this tends to the macroscopic $1/\epsilon_\infty$. 

As Y6 has such strong (high oscillator strength) and low energy excitations, the dielectric constant is far larger than typical for organic matter ($\approx 2$) and other organic electronic materials ($\approx 3$).
By the Thomas-Reiche-Kuhn f-sum rule, this is a repercussion of the total oscillator strength (a conserved quantity) being shifted down in energy, and so the screening (linked to the fastest response of the material) distance becomes shorter. 

In 1962 Penn\cite{Penn1962} provided a simple but fruitful model for the dielectric function of a semiconductor. 
This model includes the key physics of the presence of the fundamental gap, parabolic bands, and permits Umklapp scattering processes. 
A key result is that it predicts $\epsilon_\infty$ from the gap $E_g$, 
\begin{equation}
\epsilon_\infty \approx 1+\left(\frac{\hbar\omega_p}{E_g}\right)^2,
\end{equation}
where $\omega_p$ is an effective plasma frequency constrained by the Thomas-Reiche-Kuhn $f$-sum rule (oscillator-strength sum)\cite{Smith1978}, and $E_g$ is an effective interband gap.

Hence, small $E_g$ and/or large low-energy oscillator-strength density leads to a large $\epsilon_\infty$, as observed in Y6. 

Penn provides an interpolated dielectric function later reinterpreted in
a simple Lorentzian form \cite{bechstedt1992efficient}, 
\begin{equation}
    \epsilon(q)=1+\frac{\epsilon_\infty-1}{1+q^2/q_s^2},
\end{equation}

with $\epsilon_\infty$ as in the previous equation, and $q_s$ being an
effective screening wavevector, $q_s^2 = 2 \mu E_g / \hbar^2$ set to
where the electron kinetic energy starts to excite across the effective
interband gap. Clearly we go from full $\epsilon_\infty$ screening at $q=0$ to
unscreened at large $q$. 

The half-maximum midpoint, i.e. the characteristic screening distance, of this Lorentzian is at $q=q_s$. 

Now going back to the SRSH form for the range-separation, the Fourier transform of $\mathrm{erf}(\omega r)/r$ yields
a Gaussian reciprocal-space exchange fraction $1 - \exp[-q^{2}/(4\omega^{2})]$. 

The midpoint of this screening function is at $q = 2\omega\sqrt{\ln 2} \approx 1.67\,\omega$.  

Matching this to the Penn half-maximum at $q_s$ gives a direct estimate for the
range-separation parameter as 

\begin{equation}
    \omega \approx \frac{q_s}{2\sqrt{\ln 2}} \approx 0.6\, q_s.
\end{equation}

In atomic units ($\hbar=1$, distance as Bohr, energies as Hartree), we can therefore predict our range-separated parameter directly from the Penn model as

\begin{equation}
    \omega [a_0^{-1}]\approx \frac{q_s}{2\sqrt{\ln 2}} \approx 0.6\, q_s = 0.6\, \sqrt{2\mu [m_e] E_g [Ha]} .
\end{equation}

The effective mass arises from the random-phase-approximation integrating
over a parabolic-band in the Penn model. 
For organic semiconductors, this approximation is crude. 
We could probably improve this approximation with an explicit calculation of
the electronic structure, but we leave this to future work.

For low-bandgap non-fullerene acceptors such as Y6, assuming an effective mass
of 0.3 $m_e$, with a bandgap of 1.3 eV (0.05 Ha), this gives $\omega = 0.10
\  a_0^{-1}$, in striking agreement to the value from our tuning approach.

Perhaps more important than this quantitative prediction, is the  
scaling $\omega \propto \sqrt{\mu E_g}$. 
Assuming broadly similar effective mass within an organic semiconductor family,
the dominant variation is through $\sqrt{E_g}$. 
This directly explains why small-bandgap materials (i.e. almost all modern
organic semiconductors), require a smaller range-separation parameter than the
defaults tuned for typical organic molecules. 

With this information we now try and modify a popular range-separated hybrid, CAM-B3LYP.
We retain the default
exchange fractions ($\alpha = 0.19$, $\alpha + \beta = 0.65$) and change only
the range-separation parameter from $\omega = 0.33$ to $0.10\,a_0^{-1}$ (we denote this as `CAM-B3LYP-tuned' in our figures).  
Figure~\ref{tuned_dielectric} shows that the resulting excited-state properties
are in excellent agreement with the full optimally tuned SRSH functional,
despite the substantial differences in the long-range exchange fraction ($0.65$
vs.\ $1/\epsilon_\infty \approx 0.17$).  
This indicates that the key dimer photophysics requires just a reduction in the
range-separation parameter, rather than being directly affected by the
dielectric constant of the material. 
For larger assemblies, it is likely that the long-range screening (macroscopic
dielectric constant) becomes more important, and so the functional should have
a tuned $\alpha + \beta$ parameter.

\section{Conclusion}

We tuned a range-separated hybrid functional to model Y6 in the solid-state, and use this to model the excited states of the contact-pair dimers from the recent high quality crystal structure \cite{giannini2024role, Xiao2020}

Kasha's model of H (co-facial), J (head-to-tail) and V (oblique) aggregates classifies the behaviour of the contact pair dimers well, with D1, D2 and D4 exemplifying the class. 
The degree of delocalisation and FE-CT mixing varies considerable between the classes.

In the triplet manifold, for all six studied dimers, the first four excited states are Frenkel excitons (FE), followed by two charge-transfer (CT) states. 
The triplet states have modest CT-FE mixing, and have varying degrees of localisation: 
the D2 dimer states are fully delocalised; the D4 dimer states are localised to a single monomer. 
By contrast, the character of the singlet excited states are highly dimer dependent. 
For D1 and D2,  the first two states are charge-transfer in character, followed by two FE states. 
D4 demonstrates mixed CT-FE states. 

This extent of CT-FE mixing is a strong function of the dimer geometry, and we believe this plays a significant role in the strong solid-state shift of the singlet states observed in Y6. 
The larger solid-state shift in the singlet states is not just due to the additional delocalisation but also due to the larger amount of FE-CT mixing in the singlet states. 

We predict the inversion of the singlet and triplet CT states in all Y6 contact pair dimers examined.  In previous work \cite{miao2024energetic},  the singlet-triplet inversion has been proposed to reduce the triplet loss mechanism in organic photovoltaics.

We demonstrate that off-the-shelf range-separated hybrid functionals have the incorrect parametrisation for strongly-absorbing low-bandgap organic semiconductors, such as Y6. 
For Y6, this means that non-tuned range-separated functionals are less predictive than global hybrid functionals. 
Rather than an involved self-consistent fitting procedure, for CAM-B3LYP we show that a simple reduction of the range-separation parameter fixes most qualitative and quantitative errors. 

By returning to the original motivation for range-separated hybrid functionals
and comparing this to frequency-dependent screening in GW theory, and
building on Penn's\cite{Penn1962} simple semiconductor dielectric model, we
provide a direct prediction of this range-separation (screening length) in
terms of the bandgap and effective mass of the material under study.
This model suggests
that modern low-bandgap high-absorption organic
semiconductors would be well described by range-separated hybrid functions with
a corrected screening length of $\omega \approx 0.10 \  a_0^{-1}$.

\section{Author contributions}

Contributor Role Taxonomy (CRediT). 
T.W.: 
Methodology (equal);
Software (equal); 
Writing – original draft (equal); 
Writing – review and editing (equal).
I.C.: 
Formal Analysis (lead); 
Investigation (lead); 
Methodology (equal);
Software (lead);
Supervision (equal); 
Writing – original draft (equal); 
Writing – review and editing (equal).
T.R.: 
Writing - original draft (supporting); 
Methodology (supporting);
J.M.F.: 
Conceptualization (lead); 
Methodology (equal);
Supervision (lead);
Writing – original draft (equal); 
Writing – review and editing (equal).

Input files and tuning scripts implementing these workflows and analysis are available as a repository on Figshare\cite{Figshare}. 

\section{Acknowledgement}


The authors are thankful for useful discussions with Leeor Kronik, Jenny Nelson, Mariano Campoy, Gabriele D'Avino, Giacomo Londi and Samuele Giannini. 
 
J.M.F. 
is supported by a Royal Society University Research Fellowship
(URF-R1-191292). 
I.C. is supported by EPSRC (EP/Y020790/1).
T.W. and T.R. are Royal Society funded PhD students.

This work made use of the Imperial College Research Computing Service
\cite{HPC}.
Via our membership of the UK's HEC Materials Chemistry Consortium, which is funded by EPSRC (EP/R029431 and EP/X035859), this work used the \textsc{ARCHER2} UK National Supercomputing Service (http://www.archer2.ac.uk).

\bibliography{Y6-ExcitedStates}


\newpage 

\setcounter{equation}{0}
\setcounter{figure}{0}
\setcounter{table}{0}
 \setcounter{section}{0}
\setcounter{page}{1}
\makeatletter
\renewcommand{\theequation}{S\arabic{equation}}
\renewcommand{\thefigure}{S\arabic{figure}}
\renewcommand{\thetable}{S\arabic{table}}
\renewcommand{\thesection}{S\arabic{section}}
\renewcommand{\thepage}{S\arabic{page}}
\renewcommand{\bibnumfmt}[1]{[S#1]} 
\renewcommand{\citenumfont}[1]{S#1} 

\section{Basis set convergence}

\begin{figure}
    \centering
    \includegraphics[width=1\linewidth]{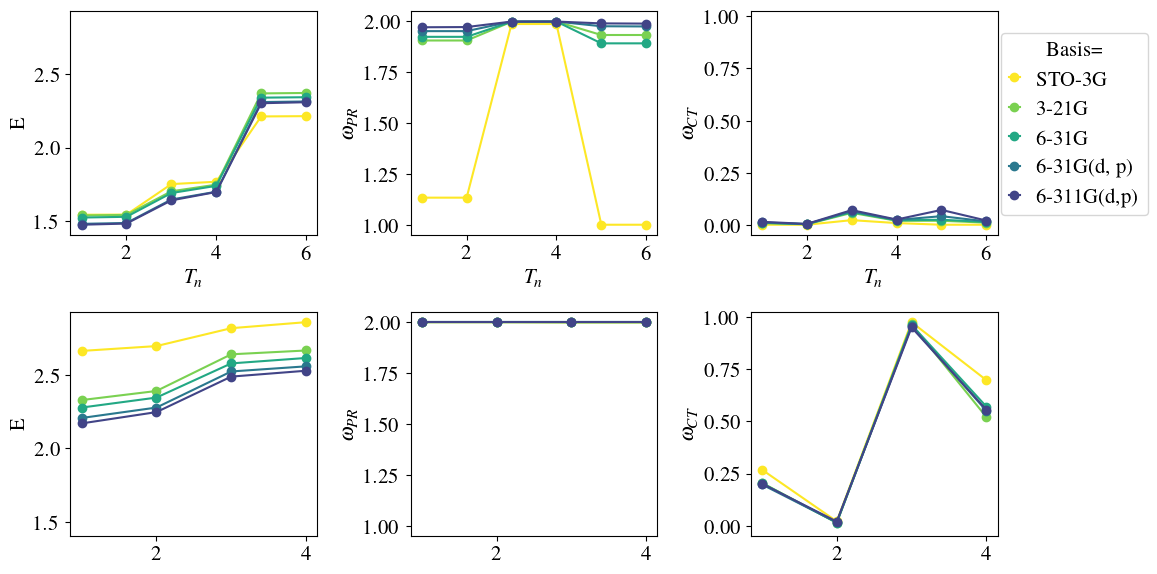}
    \caption{Figure to show how the triplet (top panels) and singlet (bottom panels) excited states of the D2 dimer using CAMB3LYP functional vary with basis set. }
    \label{converge}
\end{figure}

Before we tuned the OT-SRSH functional, we wanted to examine how the Y6 dimer singlet and triplet states converged with basis set for standard functionals. We found for all the functionals examined (B3LYP, PBE0, CAMB3LYP and wB97XD) the excited state properties were practically converged for 6-31G basis set, and so this was the basis set used to tune our OT-SRSH. This convergence for CAMB3LYP and D2 dimer is shown in figure \ref{converge}, similar behaviour is seen for other functionals and dimers.

\section{Mean squared error}

We calculate our mean-squared error as, 
\begin{equation}
\Delta_S^{TDDFT}= \sqrt{\frac{1}{6} \sum_{n=1}^{6} \left(E_{S_n}^{TDDFT}-E_{S_n}^{GW/BSE/MM}\right)^2}. 
\label{diff}
\end{equation}

\section{Effect of changing dielectric constant}

\begin{figure}
    \centering
    \includegraphics[width=1\linewidth]{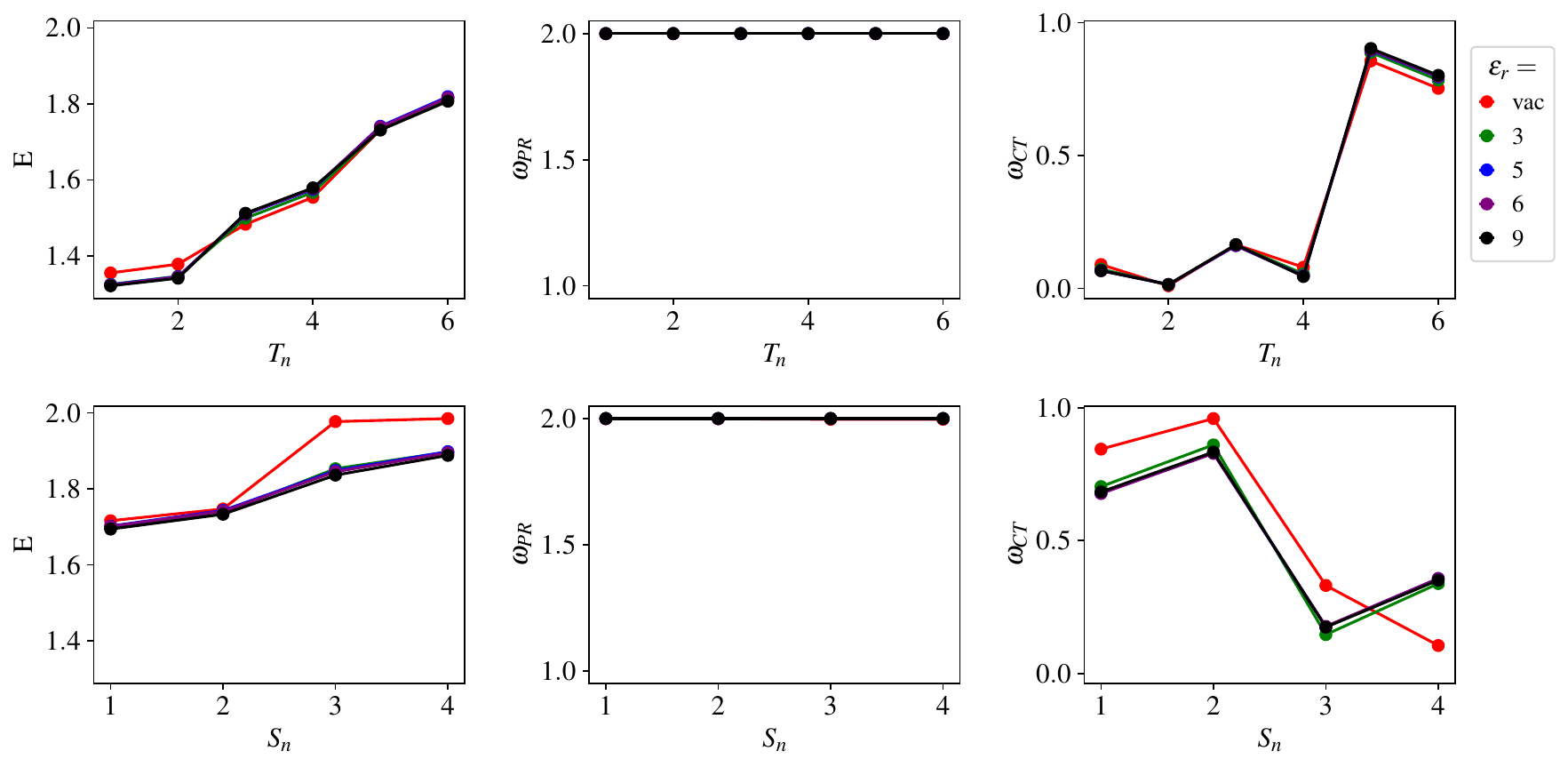}
    \caption{The properties of the singlet and triplet states of the D2 dimer depend on the dielectric constant $\epsilon_r$ used in the tuning of the OT-SRSH and in the PCM. }
    \label{epsilon}
\end{figure}

The reported experimental value of $\epsilon_r$ for Y6 films ranges from $\epsilon_r\simeq 2.5$\cite{epsilon25} to $\epsilon_r \simeq 6$\cite{epsilon573} to $\epsilon_r\simeq 11$\cite{epsilon108}.
Given the uncertainty in the literature for the values for the dielectric constant of Y6, we have examined the effect of changing the dielectric constant used in the tuning of our OT-SRSH functional and the corresponding PCM has on the properties of the singlet and triplet states, shown in figure \ref{epsilon}.

Figure \ref{epsilon} shows that within sensible range of dielectric constants of Y6, the properties of both the singlet and triplet excited states of the D2 dimer remain practically identical to those shown in the paper.

\section{Comparison of different standard functionals}

\begin{figure}[h]
    \centering
    \includegraphics[width=\linewidth]{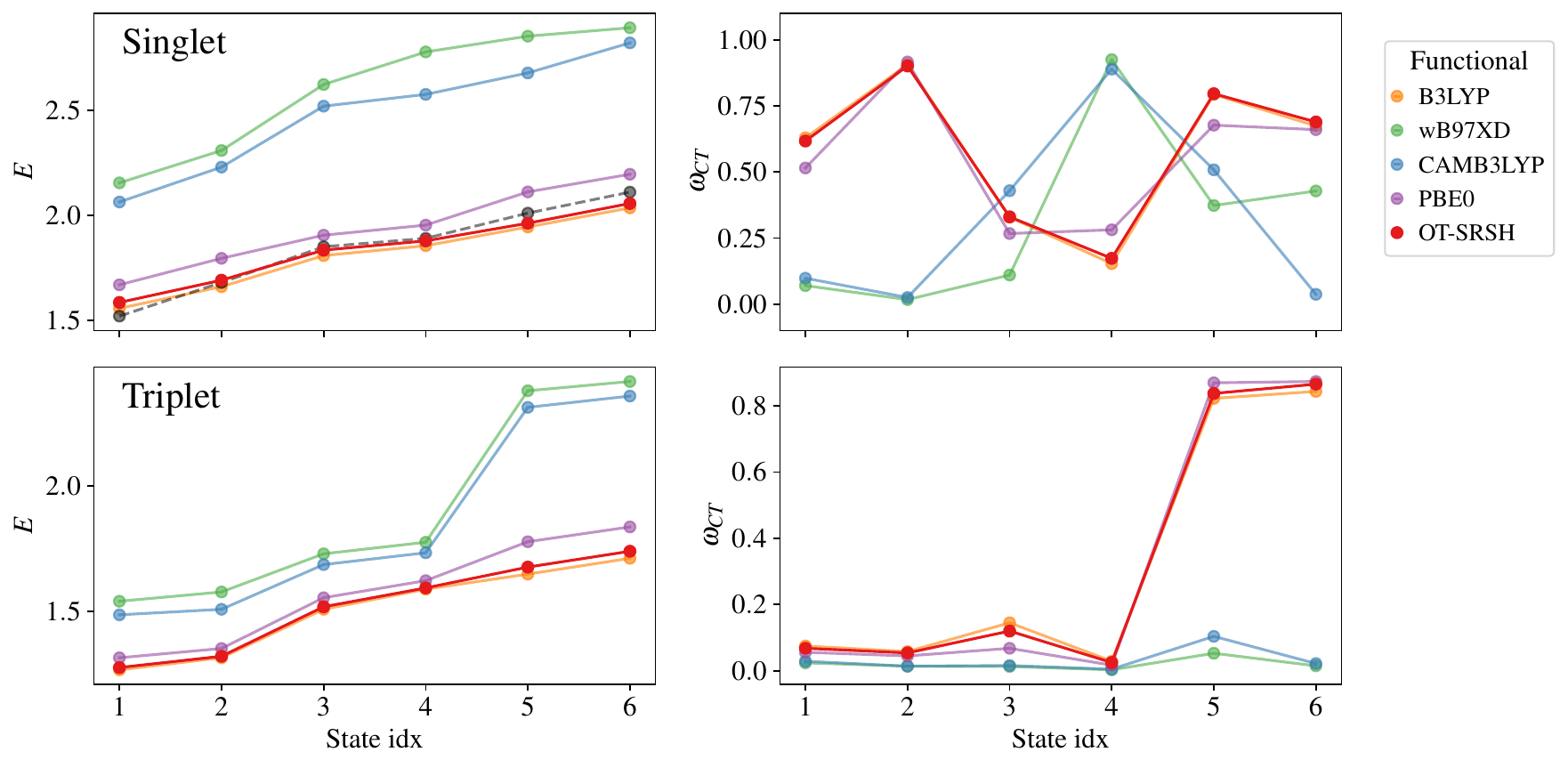}
    \caption{The energies ($E$) and the extent of charge
    transfer ($\omega_{CT}$) for the different singlet and triplet excite
    states of the D1 dimer changes with functional in PCM with $\epsilon_r=6$. 
The black dotted line is the reference GW/BSE/MM calculation\cite{akram2025analyzing}.}
    \label{D1_state}
\end{figure}

\begin{figure}[h]
    \centering
    \includegraphics[width=\linewidth]{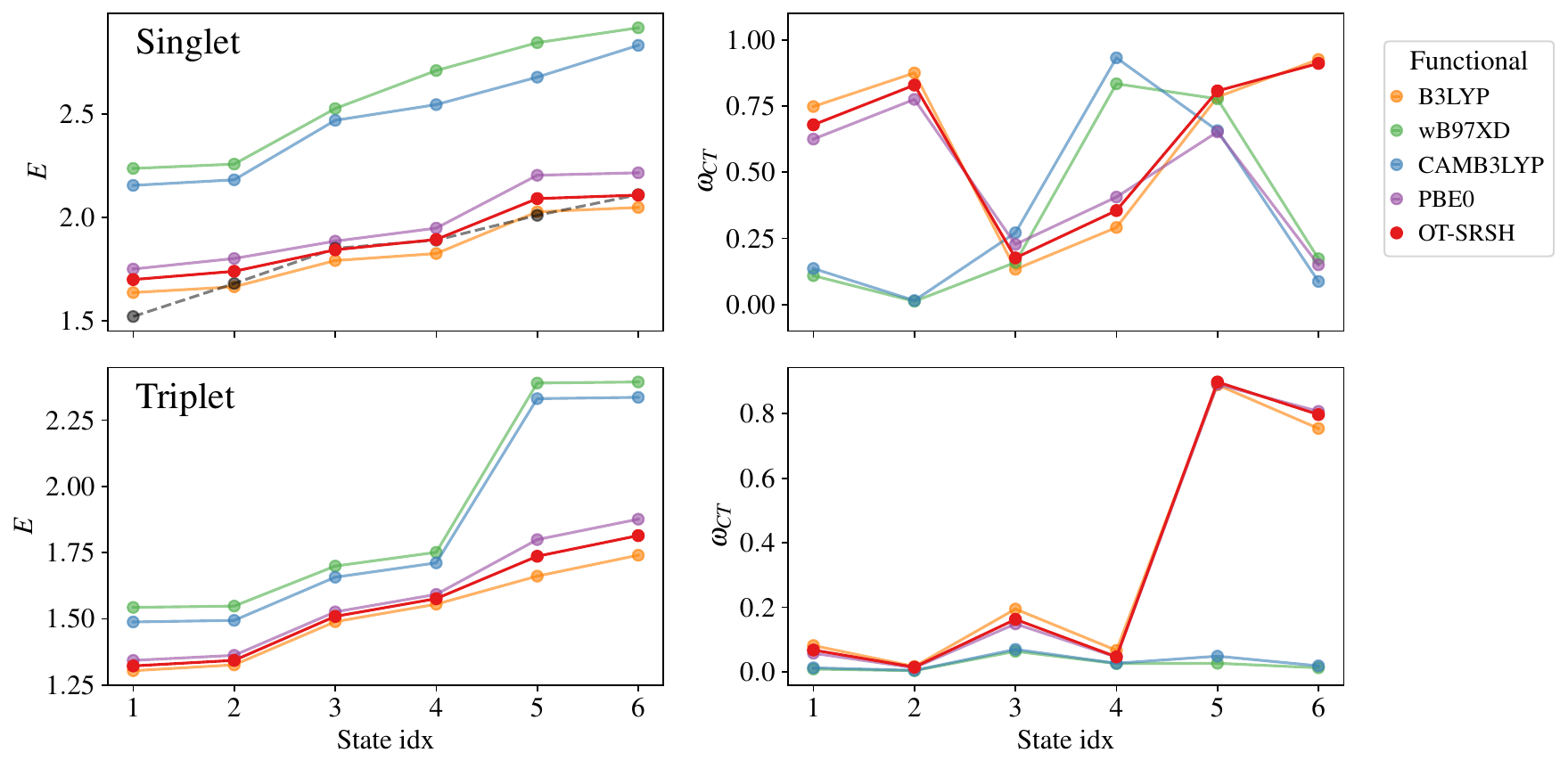}
    \caption{The energies ($E$) and the extent of charge transfer ($\omega_{CT}$) for the different singlet and triplet excite states of the D2 dimer changes with functional in PCM with $\epsilon_r=6$. 
The black dotted line is the reference GW/BSE/MM calculation\cite{akram2025analyzing}.}
    \label{D2_state}
\end{figure}

\begin{figure}[h]
    \centering
    \includegraphics[width=\linewidth]{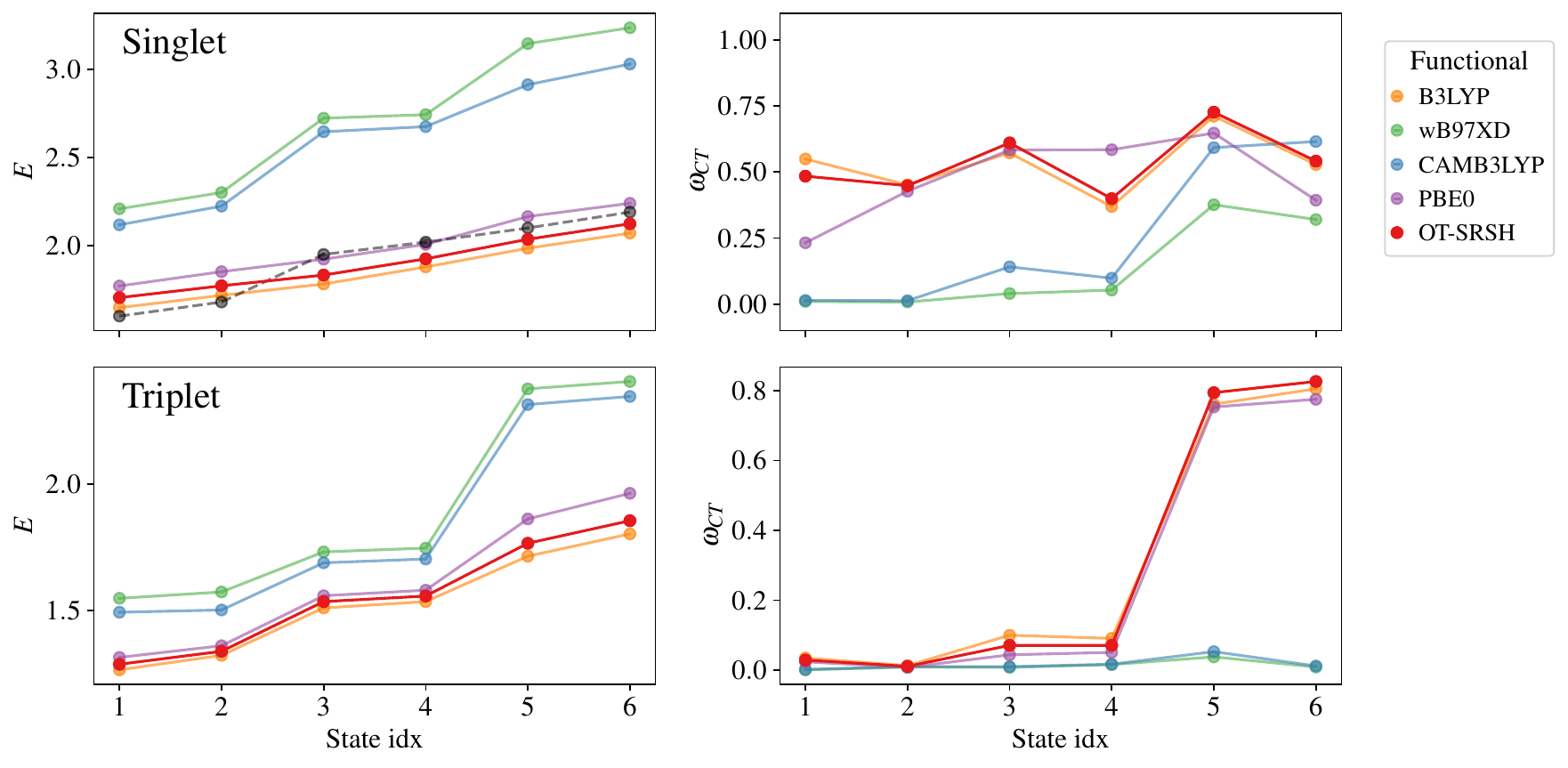}
    \caption{The energies ($E$) and the extent of charge transfer ($\omega_{CT}$) for the different singlet and triplet excite states of the D4 dimer changes with functional in PCM with $\epsilon_r=6$. 
The black dotted line is the reference GW/BSE/MM calculation\cite{akram2025analyzing}..}
    \label{D4_state}
\end{figure}

To validate our tuning, the energies of singlet excited states given by the LC-$\omega$hPBE functional were benchmarked against the results from GW/BSE/MM calculations \cite{giannini2024role}. This comparison is shown in figures \ref{D1_state}, \ref{D2_state} and \ref{D4_state} for the contact pair dimers D1, D2 and D4. These figures demonstrate good agreement between the tuned functional and the higher level GW/BSE/MM method. We find similarly good agreement results for all the other dimers.

Considering the results shown in figures \ref{D1_state}, \ref{D2_state} and \ref{D4_state} we see that the sample of functionals can be broken into two groups based on the energies and CT character of the singlet and triplet excited states in the D1, D2 and D4 dimers. 

For our OT-SRSH, together with the global PBE0 and B3LYP functionals, the lowest energy singlet excited states of the D1 and D2 dimer are 2 CT states ($\omega_{CT} \gtrsim 0.75$) followed by 2 FE states ($\omega_{CT} \lesssim 0.25$). In contrast, for CAM-B3LYP and wB97XD the lowest energy singlet excited states are 2 FE states followed by 2 CT states.  By contrast, for the D4 dimer, the OT-SRSH, B3LYP and PBE0 predict mixed CT-FE states with $\omega_{CT} \sim 0.5$ for all the four lowest energy singlet states, whilst for CAM-B3LYP and wB97XD all four lowest energy singlet states are FE states with $\omega_{CT
} \lesssim 0.25$.

A similar trend is observed for the triplet excited states of the D1, D2 and D4
dimers, where for OT-SRSH, PBE0, and B3LYP the first 4 lowest excited state are
FE states, followed by two higher in energy CT states. For CAM-B3LYP and wB97XD
the 6 lowest energy triplet states are all FE states. It is perhaps no surprise
that the hybrid functionals CAM-B3LYP and $\omega$B97XD exhibit such different
results to the OT-SRSH functional. With range-separation parameters of
$\omega=0.33 a_0^{-1}$ and $\omega=0.20 a_0^{-1}$ respectively \cite{epsilon_previous_studies}, which differ greatly to $\omega=0.11-0.13 a_0^{-1}$ for the OT-SRSH, significant difference in the electronic structure predicted by these two groups of functionals is expected.

The results of our tuning work and benchmarking demonstrate the functional dependence of both the ordering and properties of the excited states. The CAM-B3LYP functional, commonly used in the study of organic systems where CT and FE states are present, disagrees with the results from calculations using the OT-SRSH functional, and the energies given by GW/BSE theory. We therefore advise caution when using the CAM-B3LYP functional in the study of Y6 electronic structure and photophysics.  
 
One noticeable result from this work is the similarity between the results for the OT-SRSH functional and B3LYP. This may be cause for initial concern, since B3LYP is known to underestimate the energies of the CT states\cite{bogo2024benchmarking, jeon2024exploring, lan2024correction}, particularly on dissociation of the dimer \cite{magyar2007dependence}. 
However, we shall see in a later section that our OT-SRSH does not suffer from same issues inherent in the B3LYP functional. 
These results provide evidence that our optimally tuned SRSH correctly captures
the ordering of the CT and FE states, and are robust to different morphologies, whereas it is a happy coincidence that B3LYP captures the correct physics at the dimer separation.

\subsection{Size Extensivity of OT-SRSH State} 

For the OT-SRSH to be used in the study of Y6 solid and thin films, the tuning parameters must be size extensive, i.e.\ the parameters should provide consistently accurate excited state results when applied to larger aggregates. This is key to avoid the need for repeating the tuning process on various sized molecular aggregates that are used as surrogates for the solid in the `thermodynamic limit'.  

\begin{figure}
    \centering
    \includegraphics[width=1\linewidth]{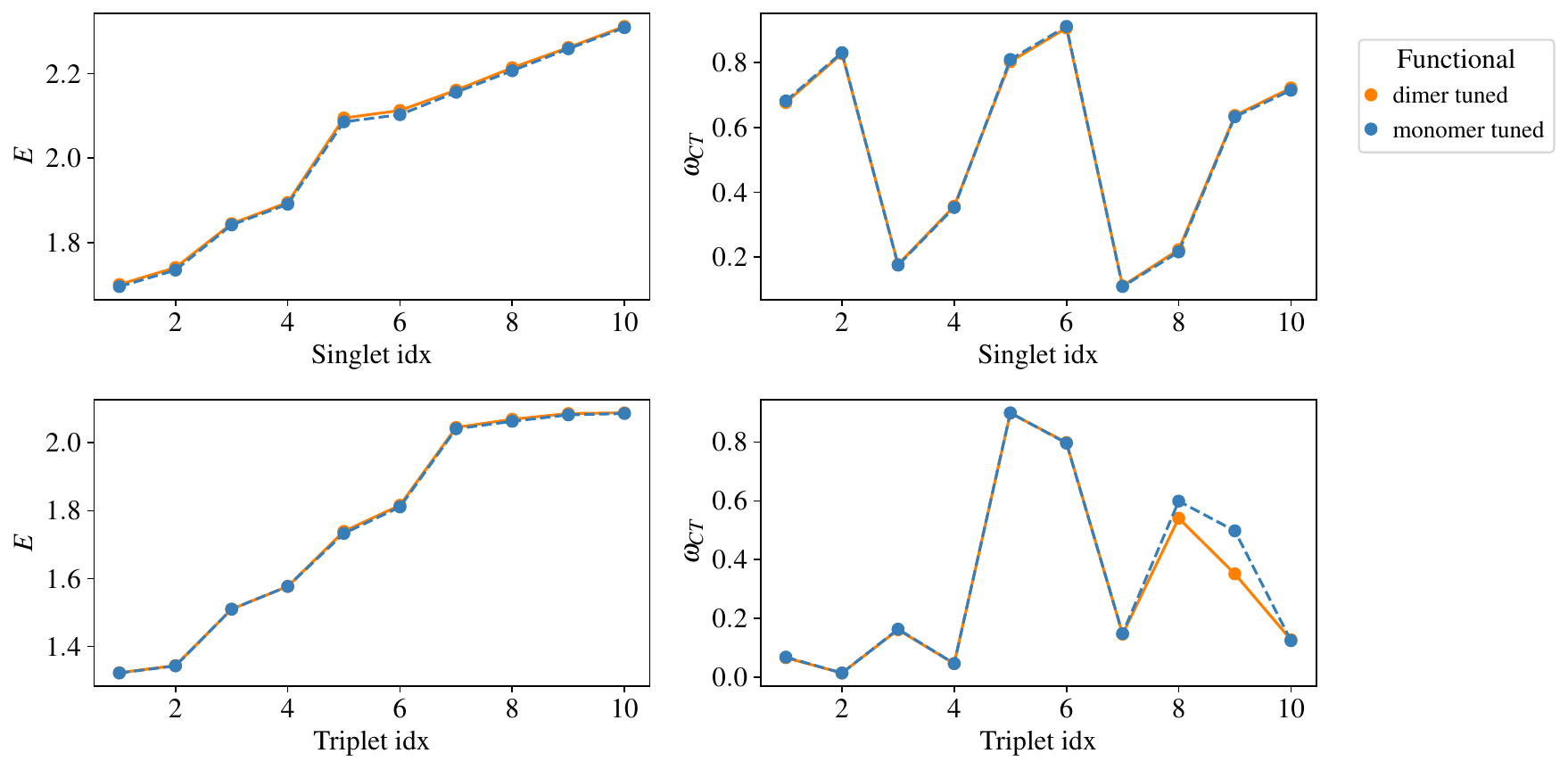}
    \caption{The singlet and triplet excited state  energies ($E$) and extent of charge transfer ($\omega_{CT}$) of the D2 dimer, where the OT-SRSH parameters used in the calculation were obtained by tuning on the monomer (blue) and the dimer D2 (orange) respectively.}
    \label{D2_size_extensivity}
\end{figure}

To demonstrate the size extensivity of the tuned functional, the excited state
properties for dimer D2 are compared for when a) calculations are run using the
tuned parameters for dimer D2, and b) calculations are run using parameters
tuned for the monomer. The results of this comparison (Fig.\
\ref{D2_size_extensivity}) show that for both singlet and triplet excited
states, the state energies and the extent of charge transfer ($\omega_{CT}$)
are independent of whether the tuning was done on the monomer or dimer. This
demonstrates the size extensivity of the tuned parameters. Additionally, we
find that for OT-SRSH tuned on both the monomer and the dimer, the singlet and
triplet excited states in dimer D2 have $\omega_{PR} \simeq 2$, meaning the
excited states are completely delocalised across both monomer units. We find
similar results for all the contact-pair dimers.  Therefore, we expect that for calculations on Y6 molecular aggregates, the OT-SRSH parameters tuned on the monomer can be applied to larger molecular aggregates. 

A similar result using different analysis was found in the recent paper by Akram et al.\cite{akram2025analyzing}.

\subsection{Properties of States with Increased Dimer Separation}

\begin{figure}[h]
    \centering
    \includegraphics[width=1\linewidth]{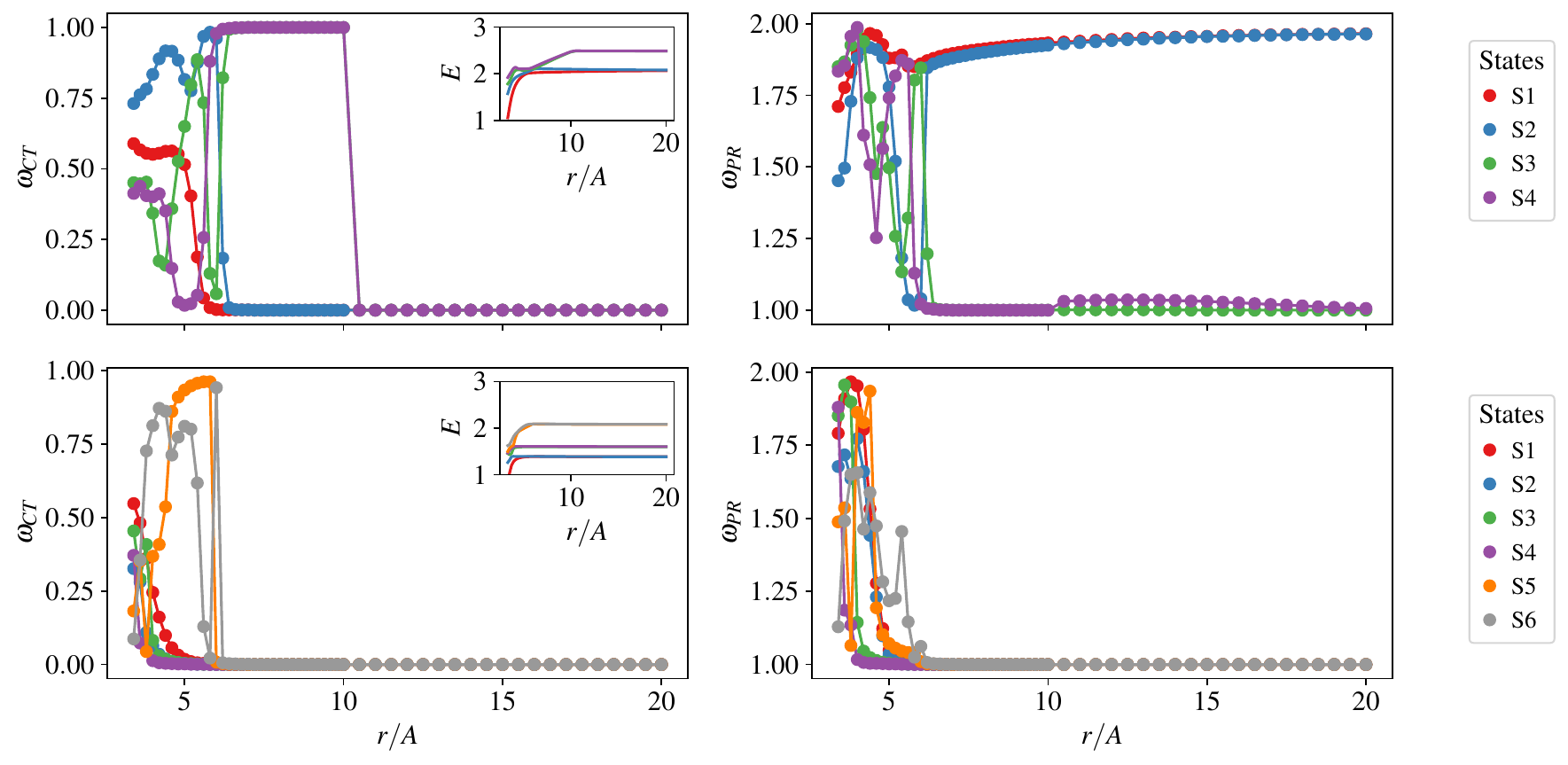}
    \caption{The energy ($E$), CT character ($\omega_{CT}$) and extent of delocalisation ($\omega_{PR}$) of the singlet (top panels) and triplet (bottom panels) excited states of the Y6 dimer change with the separation between the monomer units for our OT-SRSH functional. 
}
    \label{dimer_seperation_OT}
\end{figure}

\begin{figure}
    \centering
    \includegraphics[width=1\linewidth]{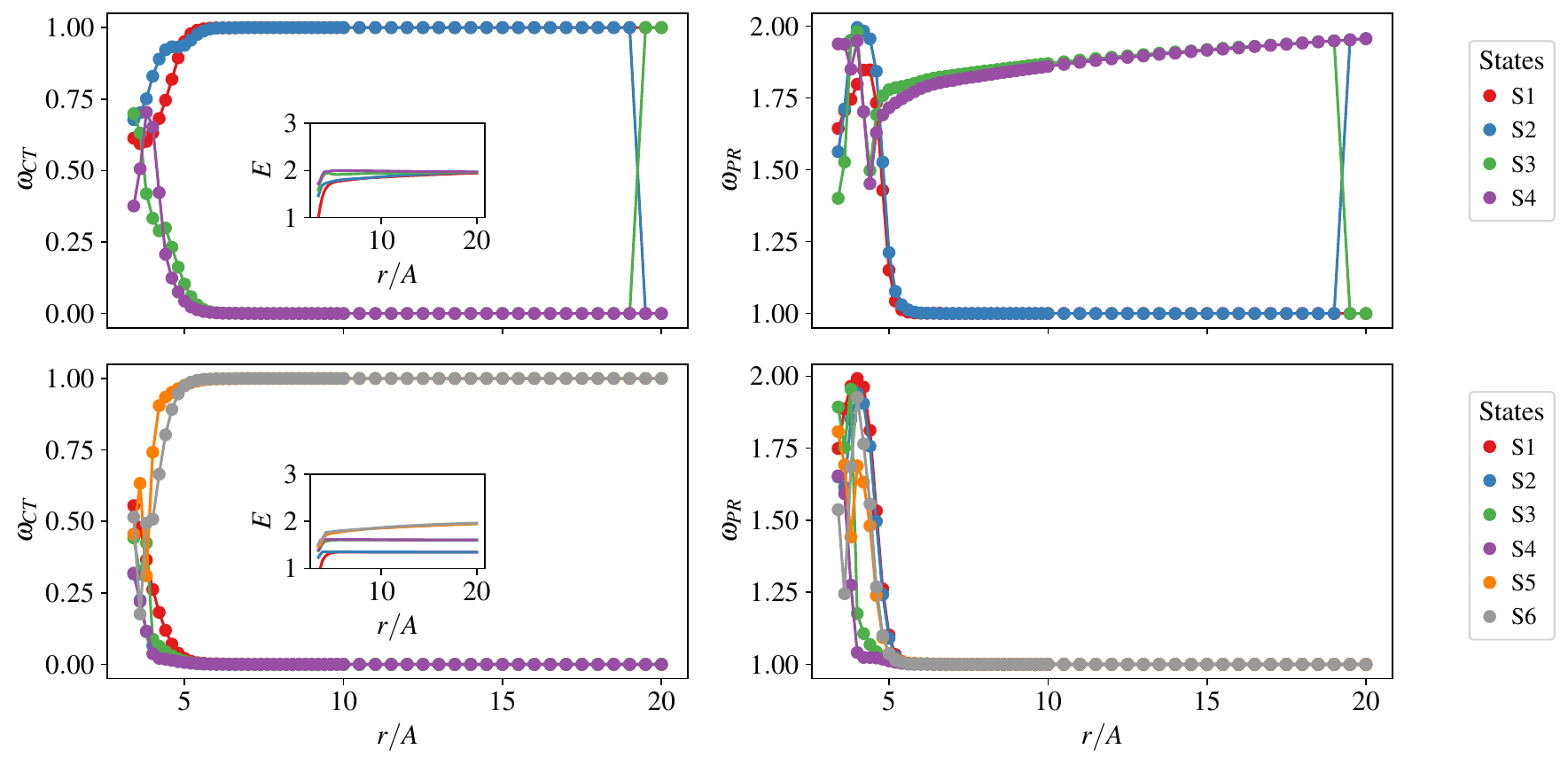}
    \caption{The energy ($E$), CT character ($\omega_{CT}$) and extent of delocalisation ($\omega_{PR}$) of the singlet (top panels) and triplet (bottom panels) excited states of the Y6 dimer change with the separation between the monomer units for the B3LYP functional. }
    \label{dimer_seperation_B3LYP}
\end{figure}

In addition to size extensivity, it is important that the tuned functional is
able to capture the correct ordering of the CT and FE states and the extent of
delocalisation/size of the exciton in Y6 crystals. To probe this, we consider
how properties of the first 4 singlets and 6 triplets, extent of charge
transfer and delocalisation, change as the separation distance between the two
dimers increases, both for our OT-SRSH functional (figure
\ref{dimer_seperation_OT}) and B3LYP (figure \ref{dimer_seperation_B3LYP}).

Physically, we would expect the excited states of the dimer to localise in
separate monomer units as the dimer distance increases, with
$\omega_{PR}\rightarrow 1$ at large separations. 
We expect that this will happen faster for the triplet and CT states than for
the FE states, due to F\"orster term facilitating easier delocalisation in FE
states than other excitation\cite{olaya2011energy}.

Furthermore, for both our OT-SRSH and B3LYP, if the correct physics is captured, we would expect that whilst the charge transfer states are the lowest energy singlet states at short separations for the Y6 dimers, as the separation increases, these states would become increasingly higher in energy. 
Initially, the $S_1$ and $S_2$ states would move from being predominantly CT states to predominantly FE states, meaning the $S_3$ and $S_4$ states would move from FE states to CT states. 
As the distance between the monomers increases further, we would expect that all four singlet states would be FE exciton states. 

Similarly, for the triplet states for both our OT-SRSH and B3LYP functionals, if the correct physics is captured,  at short separations we expect from our previous calculations that first four excited states are all FE states, followed by two higher in energy CT states, and we would expect that as monomer separation increased eventually all triplet states would be FE states.

\begin{figure}
    \centering
    \includegraphics[width=1\linewidth]{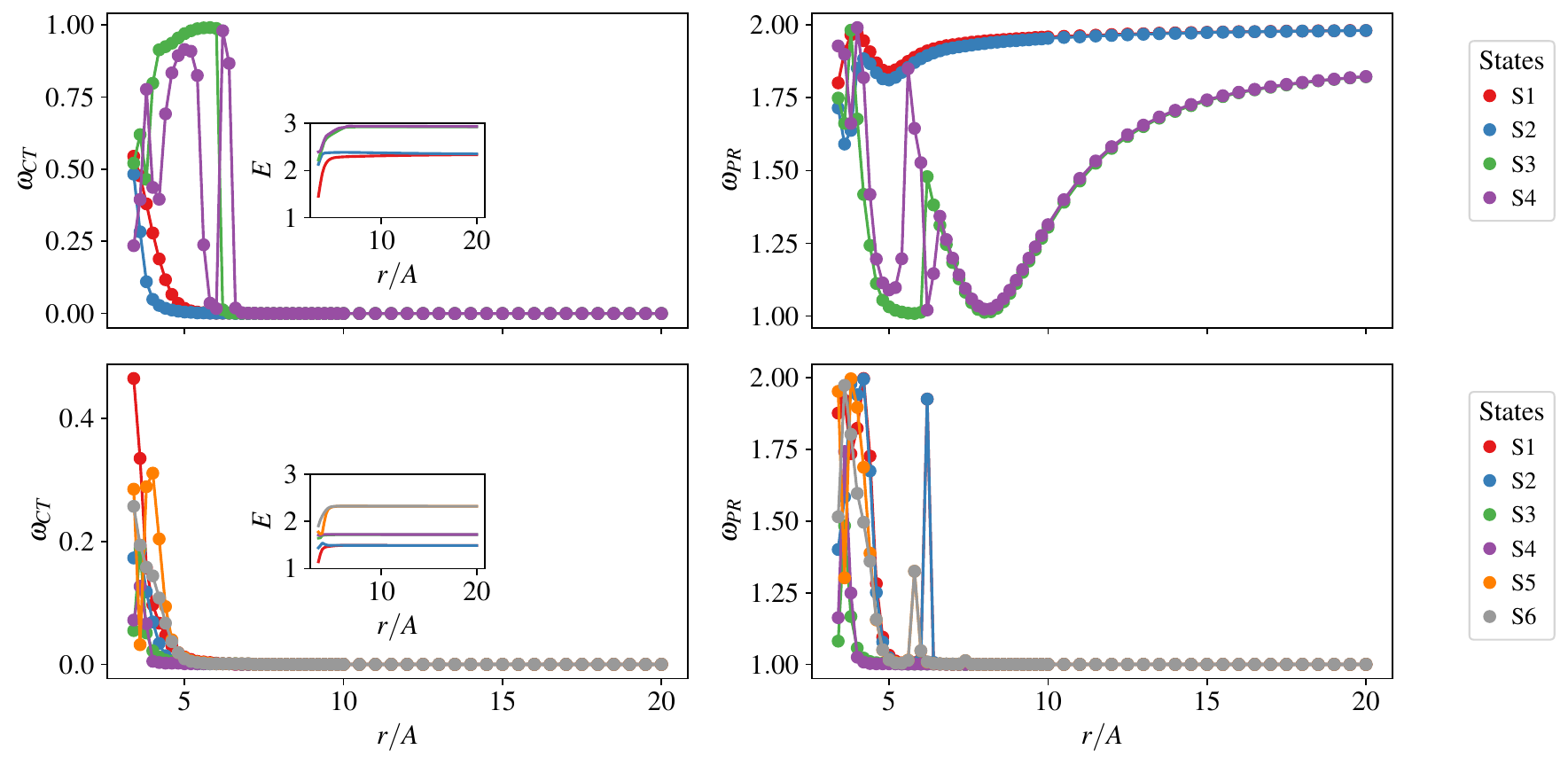}
    \caption{The energy ($E$), CT character ($\omega_{CT}$) and extent of delocalisation ($\omega_{PR}$) of the singlet (top panels) and triplet (bottom panels) excited states of the Y6 dimer change with the separation between the monomer units for the CAM-B3LYP functional.}
    \label{dimer_seperation_CAM-B3LYP}
\end{figure}

Fig.\ \ref{dimer_seperation_OT} shows that the tuned OT-SRSH functional is able
to describe the behaviour of the CT character of the singlet and triplet states
($\omega_{CT}$) as the monomer separation is increased. In contrast, figure
\ref{dimer_seperation_B3LYP} shows that the B3LYP functional is unable to fully
describe this behaviour.  In this case, the lowest two singlet states remain CT
states even at a monomer separation of 20 Å which is not physical. This
leads us to caution the use of B3LYP functional in examining the CT character of larger size aggregates of Y6 or other OPV systems. It is, however, clear that our OT-SRSH could be used to look at CT states in larger aggregates.

We now consider whether any of the functionals considered in this paper are
able to describe the physics of FE states at large separations. Figures
\ref{dimer_seperation_OT}, \ref{dimer_seperation_B3LYP} and
\ref{dimer_seperation_CAM-B3LYP}  show that at large separations for OT-SRSH,
B3LYP and CAM-B3LYP functional, some of the FE singlet states of the D2 dimer
remain delocalised even at unphysical separations of 20 Å. Further work is
therefore clearly needed to develop a functional capable of capturing both the
CT character of the states and delocalisation of the singlet states, and we
caution against using the OT-SRSH functional when examining the FE exciton 
size of Y6 molecular aggregates.

\subsection{Reorganisation Energy}
When a molecule undergoes an electronic transition, the electron density changes, so that the equilibrium geometry in the new electronic state usually differs from the equilibrium geometry in the initial state, reflecting the change in the electrostatic distribution between electrons and nuclei. 
The energy associated with the rearrangement of nuclei as a result of an
electronic transition is called the reorganisation energy and it plays a crucial role in enhancing efficiencies in organic semiconductors \cite{Azzouzi2018,Wang2011,Lin2019}. 

Several computational approaches can be used to determine the reorganisation energy, which is conventionally partitioned into an intramolecular component (inner reorganisation energy) and an extramolecular component (outer reorganisation energy). 
Our focus will be solely on the former. 
The most common technique for calculating the total intramolecular
reorganisation energy is Nelsen's 4-Point (4P) method \cite{Nelsen1987}, which
requires four single-point energy calculations performed at the equilibrium
geometries of the two electronic states and their respective Franck-Condon
geometries, accounting for both forward and backward transitions. However, as
the 4P method only yields the total value, an alternative approach developed by
J. R. Reimers \cite{Reimers2001} is often utilised to resolve the individual
vibronic coupling strength contributed by each normal mode to the overall
reorganisation energy.

\begin{figure}
    \centering
    \includegraphics[width=0.75\linewidth]{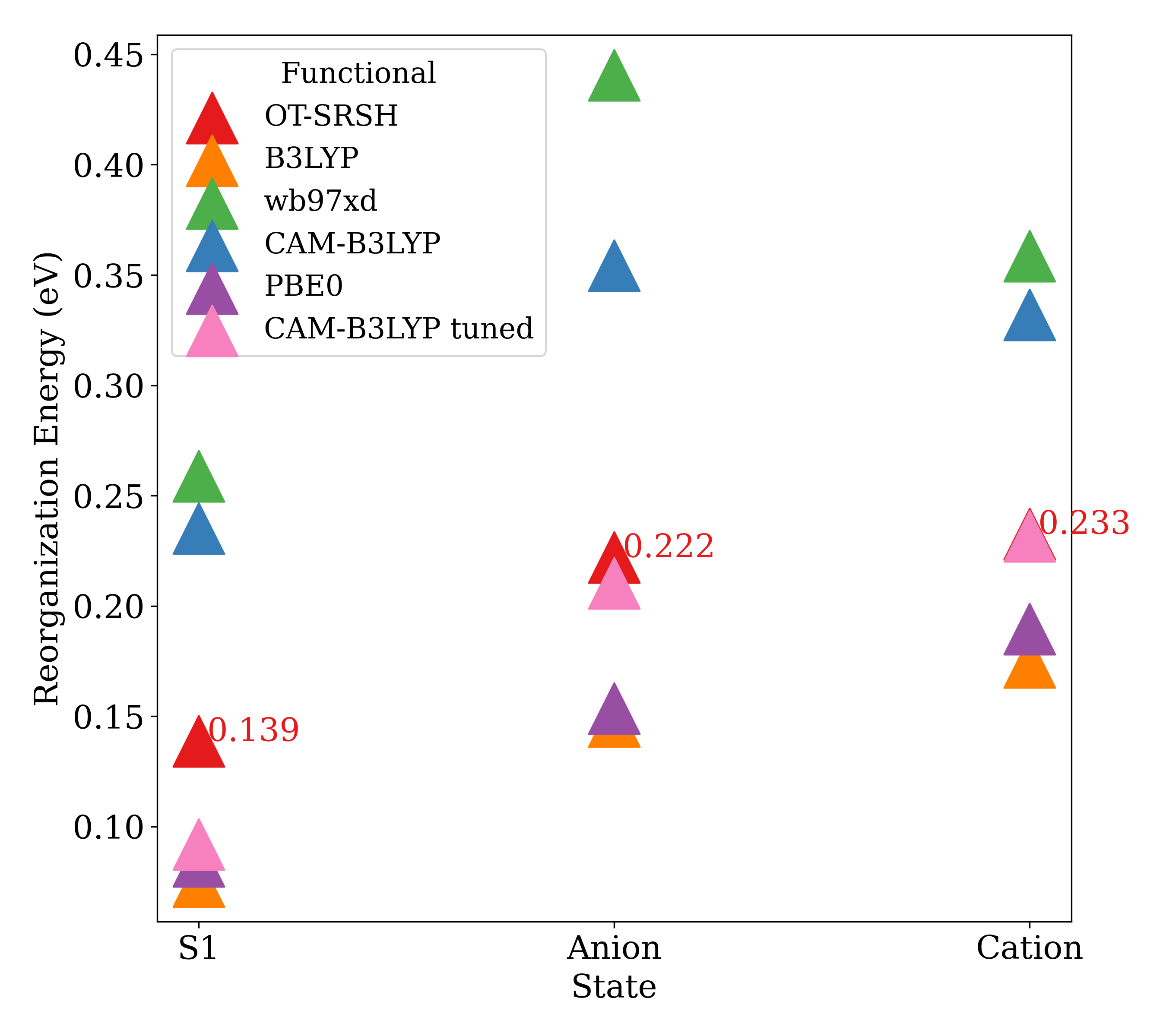}
    \caption{Reorganisation energies for the first excited state of the different functionals for the Y6 monomer in vacuum, as well as for the anion and cation. Values were obtained via the 4P-approach.}
    \label{fig:4P_RE_Excited_Charged_States}
\end{figure}

To validate the tuning approach and to ensure that it yields sensible
reorganisation energy values, we first calculated the intramolecular S1 excited
state reorganisation energy, alongside the hole and electron reorganisation energies (c. fig. \ref{fig:4P_RE_Excited_Charged_States}), for the Y6 monomer in vacuum using the 4P approach. 
The overall trend confirms that Y6 exhibits a very low excited state
reorganisation energy for S1, which notably lies below both its hole and
electron reorganisation energies. 
Furthermore, these results demonstrate for the OT-SRSH reasonable agreement with experimentally determined values, which typically fall around 85 meV \cite{Hanbo2026}  - 103 meV \cite{Kashani2023}. 
Across the excited and charged states, the reorganisation energies predicted by B3LYP, PBE0, and tuned CAM-B3LYP are largely consistent with the OT-SRSH results. In contrast, $\omega$B97X-D and standard CAM-B3LYP yield distinct absolute values, forming a separate cluster. 
To further test the tuning, it is instructive to examine the normal mode
contributions to the reorganisation energy using the method developed by Reimers \cite{Reimers2001}. 
Since the normal modes are fundamentally derived from the diagonalisation of
the Hessian matrix, analysing their individual contributions allows us to gain deeper insight into the underlying electronic structure calculated by the different functionals, moving beyond an assessment based solely on total energy values. 

\begin{figure}
    \centering
    \includegraphics[width=0.75\linewidth]{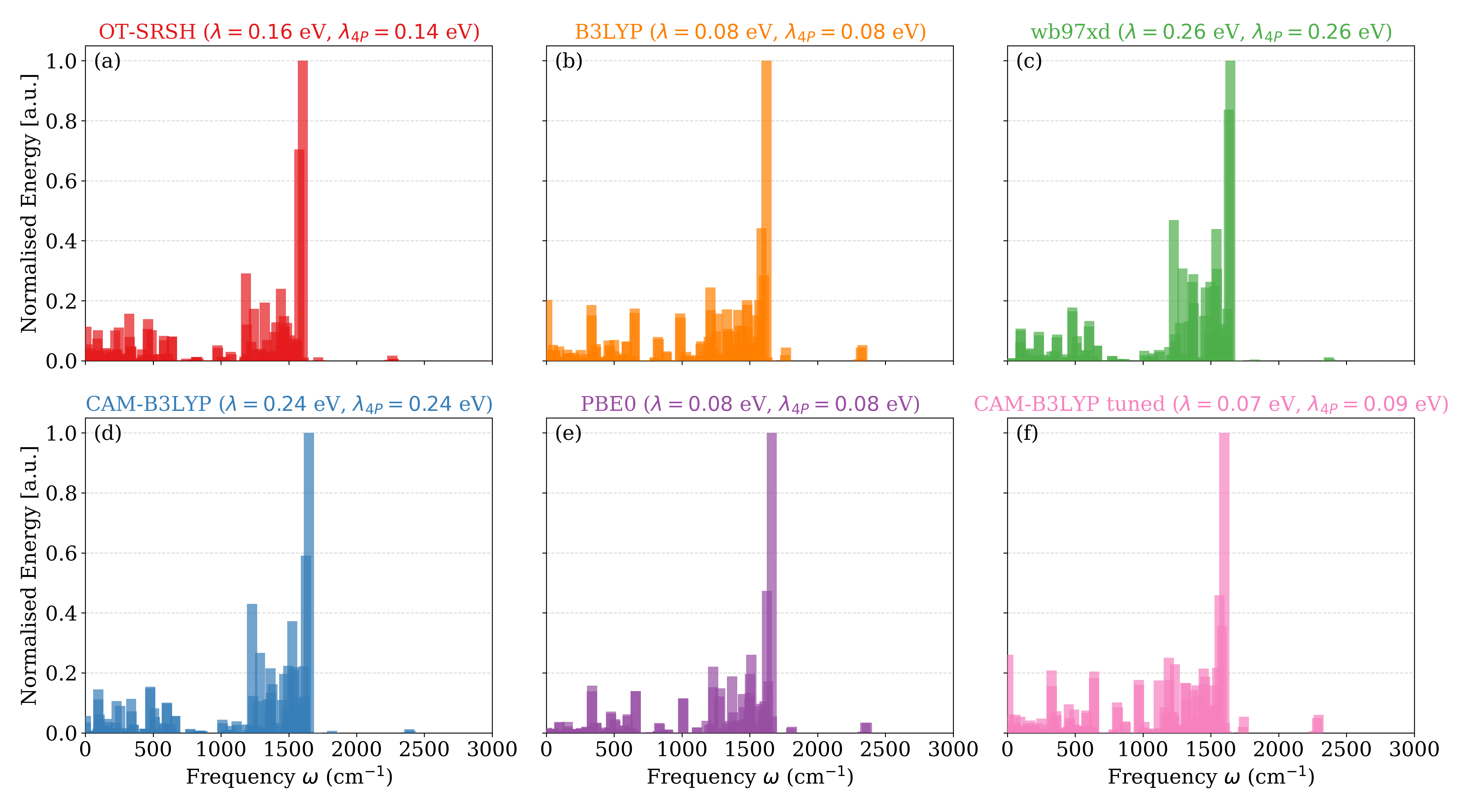}
    \caption{Normal mode contribution to the overall reorganisation energy across different functionals. For each functional the values are normalised to the mode that couples the strongest.}
    \label{fig:dushin_spectra_normalised}
\end{figure}

Figure \ref{fig:dushin_spectra_normalised} presents the calculated normal mode
contributions to the reorganisation energy using the DUSHIN code by Reimers' which are normalised to the mode value that couples the strongest. 
The close agreement between the total reorganisation energies obtained by the 4P approach and by summing the contributions of all normal modes across the different functionals validates the use of the harmonic approximation in these calculations. 
Furthermore, all functionals consistently identify the primary vibrational contribution occurring near $1600 \text{ cm}^{-1}$, which corresponds to the breathing mode of the $DA^{'}D$ structure. Interestingly, the vibrational mode around $1350 \text{ cm}^{-1}$, primarily attributed to stretching motions within the acceptor regions, exhibits a significant relative contribution for the $\omega\text{B97XD}$ and $\text{CAM-B3LYP}$ functionals, registering as the second largest component. In contrast, this mode's contribution is notably smaller when calculated with the other functionals.

\section{Dimer aggregate properties}

\begin{figure*}
    \centering
    \includegraphics[width=1\linewidth]{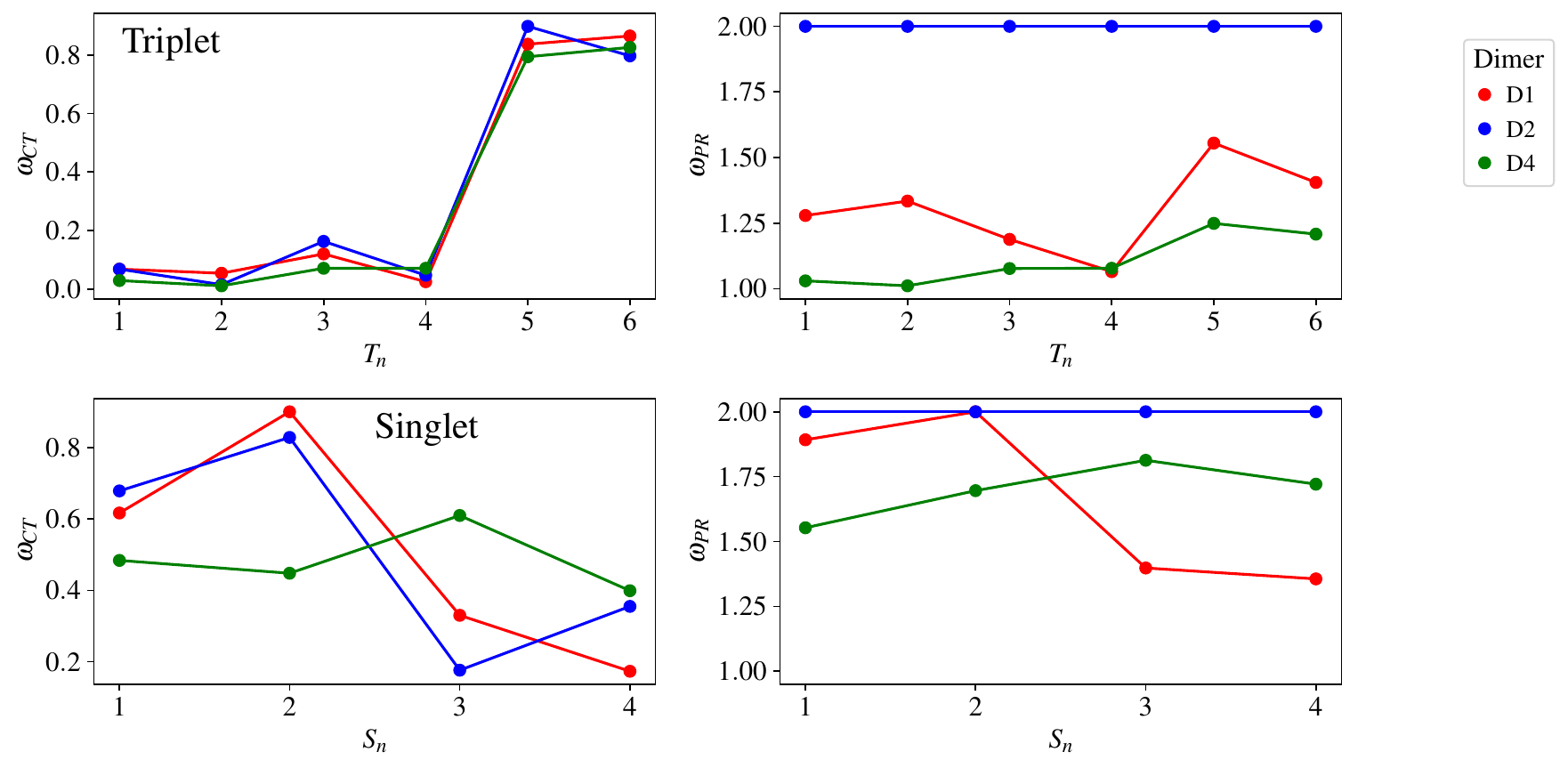}
    \caption{The extent of charge transfer ($\omega_{CT}$) and extent of delocalisation ($\omega_{PR}$) of the singlet and triplet states of the D1, D2 and D4 dimer.}
    \label{excited_properties}
\end{figure*}

\begin{figure*}
    \centering
    \includegraphics[width=1\linewidth]{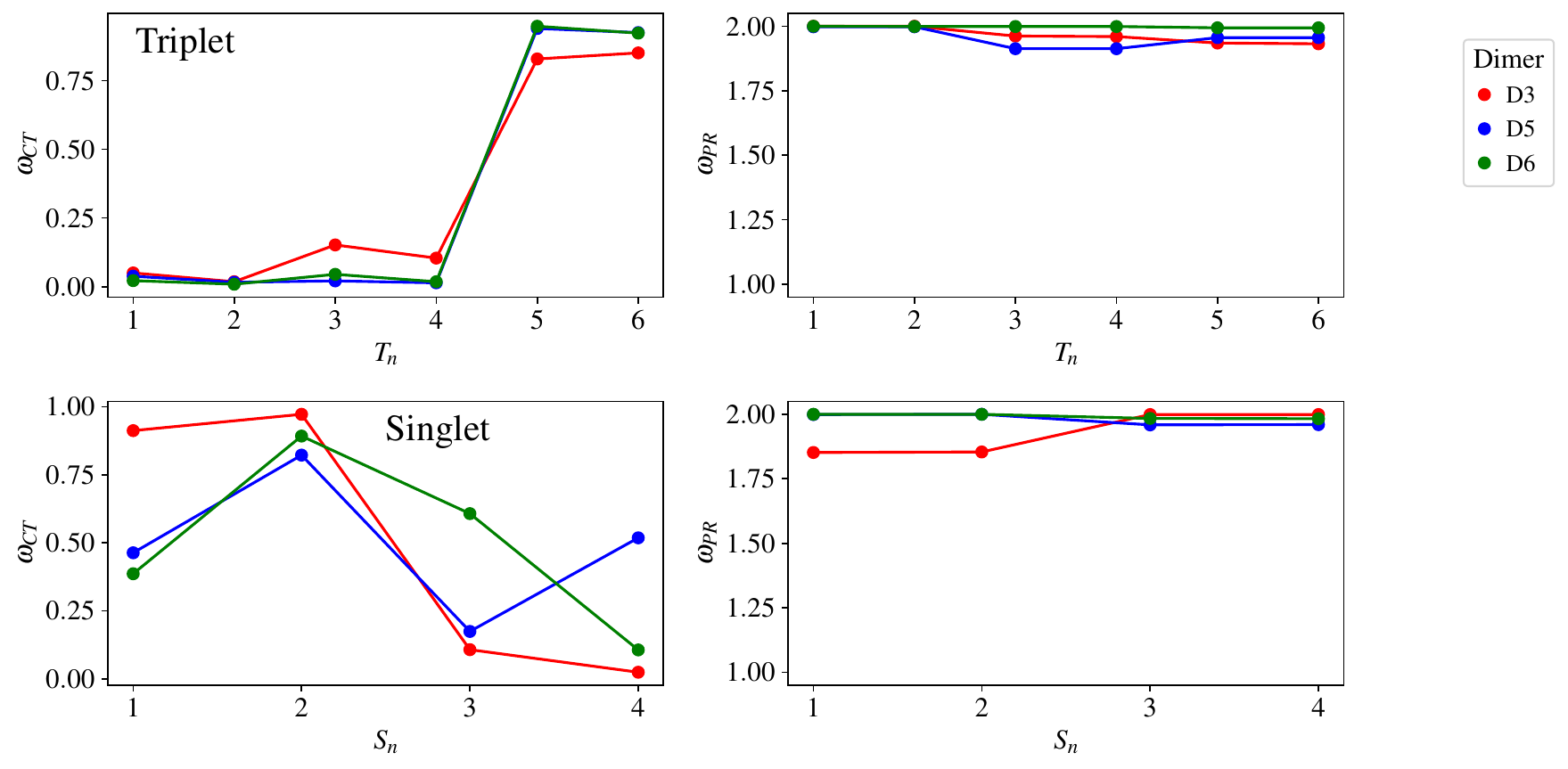}
 \caption{The extent of charge transfer ($\omega_{CT}$) and extent of delocalisation ($\omega_{PR}$) of the singlet and triplet states of the D3 (H aggregate), D5 (J aggregate) and D6 (J aggregate).}
    \label{excited_properties_2}
\end{figure*}

In Figs.\ \ref{excited_properties} and \ref{excited_properties_2} we show the
behaviour of the different contact pair dimers in the Y6 crystal.  In work by
Giannini \textit{et al.\ }\cite{giannini2024role}  dimers D1 and D4 are
considered H aggregates, and D2 is a J aggregate. This classification is based
on Kasha's model, where only the relative orientation of the monomer dipole
moments is considered.  The D4 dimer is distinct from the other contact dimers,
as it has little spatial overlap of the monomers. 
In previous work \cite{Han2020} this type of dimer has been classified as a V aggregate. 
In the case of the contact dimers of Y6 our analysis below shows this difference in classification is warranted. 

For all six different types of dimers considered in Figs.\ \ref{excited_properties} and \ref{excited_properties_2}, we observe that the four lowest energy triplet states are all FE states ($\omega_{CT}<0.25$) followed by 2 higher in energy CT states ($\omega_{CT}>0.75)$, with a small amount of CT-FE mixing which slightly depends on the dimer considered. The extent to which the triplet states are delocalised across the dimer ($\omega_{PR}$) depends strongly on the dimer considered. The triplet states in e.g. dimer D2 are entirely delocalised across both monomer units with $\omega_{PR} \simeq 2$, while the dimer D4 triplet states are nearly completely localised on a single monomer unit, $\omega_{PR} \simeq 1$.

In contrast, Figs.\ \ref{excited_properties} and \ref{excited_properties_2} show that for dimers D1, D2, D3, D5 and D6 the two lowest energy singlet states are CT states, followed by two higher in energy FE states. For the dimer D4 all four lowest energy states are essentially mixed CT-FE states. The OT-SRSH functional also predicts that the degree of CT-FE mixing is greater for all singlet states than for the triplet states.  For all the dimers considered the four lowest energy states are significantly delocalised across both monomer units, with the singlet states on dimer D2 being completely delocalised with $\omega_{PR} \simeq 2$, while the states on dimer D4 are more localised with $\omega_{PR} \simeq 1.6$. Additionally, the singlet states for some dimers are more delocalised than their triplet counterparts. 

We consider the more significant difference between the singlet and triplet states to be the extent of CT-FE mixing between states. This clearly plays a role in the larger shift  in energy of the singlet than triplet states. This goes beyond the view that this difference is solely due to the difference in the singlet and triplet extent of delocalisation. To further illustrate this, we consider the D2 dimer: both the singlet and triplet states are fully delocalised across the dimer. The cause of the larger gap in energy of the two lowest energy FE singlet states than the lowest energy FE triplet states can therefore only be due to the difference in CT-FE mixing.

To explain some of the different behaviour of the dimers, we considered the correlation between the electron and hole of a given exciton on fragments of the dimer. 
The structures of the different dimers (shown in figs. \ref{D1_dimer}, \ref{D2_dimer} and \ref{D4_dimer}) show the H (co-facial), J (head-tail) and V (oblique) orientations. 
There is significant wavefunction overlap in dimers D1 and D2, but limited overlap for dimer D4.

Fig.\ \ref{excited_properties} shows that when there is overlap between the two core (donor) units, that the $S_1$ state is a CT state, whilst the $S_3$ state is an FE state. However, when there is limited overlap between the two donor units, as in dimer D4, Fig.\ \ref{excited_properties} shows that the 4 lowest energy singlet states are mixed FE-CT states. 

\begin{figure}
    \centering
    \includegraphics[width=1\linewidth]{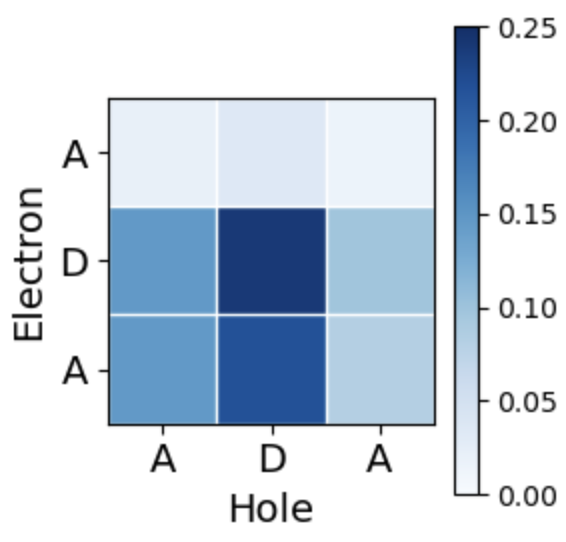}
    \caption{Electron/Hole correlation plot of the $S_1$ state for the Y6 monomer. The monomer is fragmented into two end acceptor units and a central donor unit. The fragmentation scheme is shown in Fig.\ \eqref{monomer fragmentation}. The colourbar corresponds to the 1 particle transition density (given in terms of the electron and hole position) for the given excited state. On-diagonal elements correspond to the electron and hole being correlated on the same fragment, e.g.\ a Frenkel Exciton and off-diagonal elements correspond to Charge-Transfer excitation. }
    \label{monomer electron hole correlation}
\end{figure}

To understand this, we first consider the monomer $S_1$ state. Shown in Fig.\ \ref{monomer electron hole correlation}, the predominant contributions of the S1 state have FE character, with the electron and hole both localised on core (donor) unit. There are minor CT contributions from when either the electron or hole are localised on the core or end units respectively. 


In dimer D2, where there is overlap between the two core units, Fig.\ \ref{e/h
correlation} shows that $S_3$, an FE state, is a delocalised state. The main FE
contributions are observed on the core of each monomer. The contributions with
CT character are minor, and occur with the electron and hole on the same
monomer. By contrast, the main contribution for the CT state, $S_1$,  on dimer
D2 arises from when the electron and hole are on core units of different
monomers.   We, therefore, see that when there is overlap between the D units,
the electronic structure of the FE and CT states are essentially a delocalised
version of the electronic structure of the monomer, and a charge-separated
version of the electronic structure of the monomer, facilitated by significant overlap. 
Here, the unique properties of Y6 mean the charge-transfer state is lower in energy than the FE state. 

However, in dimer D4, where there is limited overlap between the two core units Fig.\ \ref{excited_properties} shows that all 4 lowest energy singlet states are mixed FE-CT states. In dimer D4, contact between the monomers is exclusively between two end (acceptor) units ($A^{\prime}1$ and $A^{\prime}2$). For parts of the monomer wavefunction when both electron and hole are on donor unit $D2$ or when either the electron or hole are on $D2$ and the other is on the acceptor $A2$ there is limited delocalisation of the wavefunction in the dimer, as these regions of the wavefunctions in dimer D2 do not overlap significantly with the wavefunction of dimer D1.

By contrast, when either the electron or hole are on $D2$ and the other is on
the acceptor $A^{\prime}2$, there is overlap of wavefunction of dimer D1
facilitating the formation of a CT like state. 
This means that for D4 the S1 state is a mixture of localised FE state on D2 and a CT state from electron or hole transfer across from the $A^{\prime}1$  to $A^{\prime}2$ fragments. Since the S1 and S2 state are mixed FE-CT states, the S3 and S4 states also have mixed character.

\begin{figure}
    \centering
    \includegraphics[width=1\linewidth]{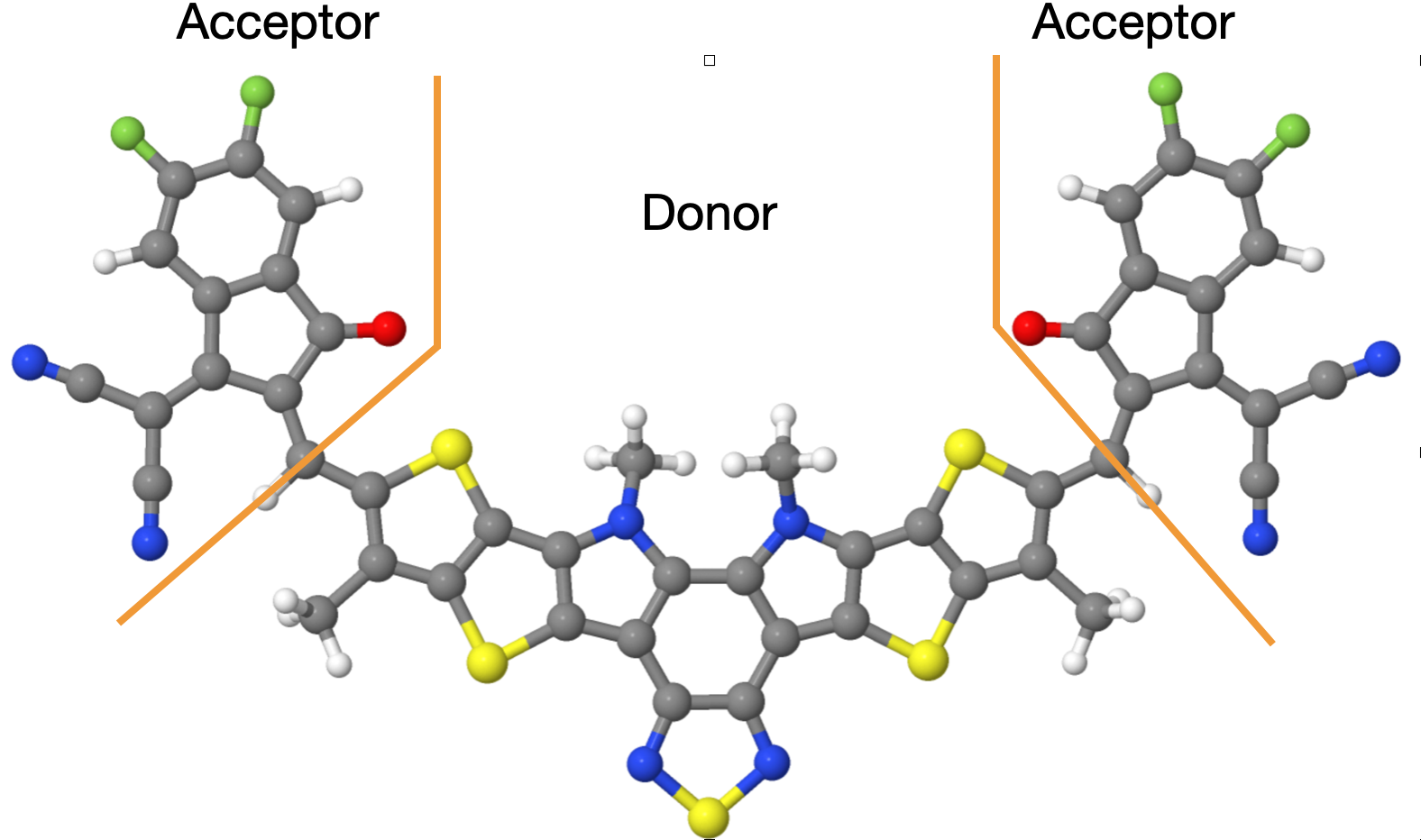}
    \caption{Fragmentation scheme of the Y6 monomer.}
    \label{monomer fragmentation}
\end{figure}
\begin{figure}
    \centering
    \includegraphics[width=1\linewidth]{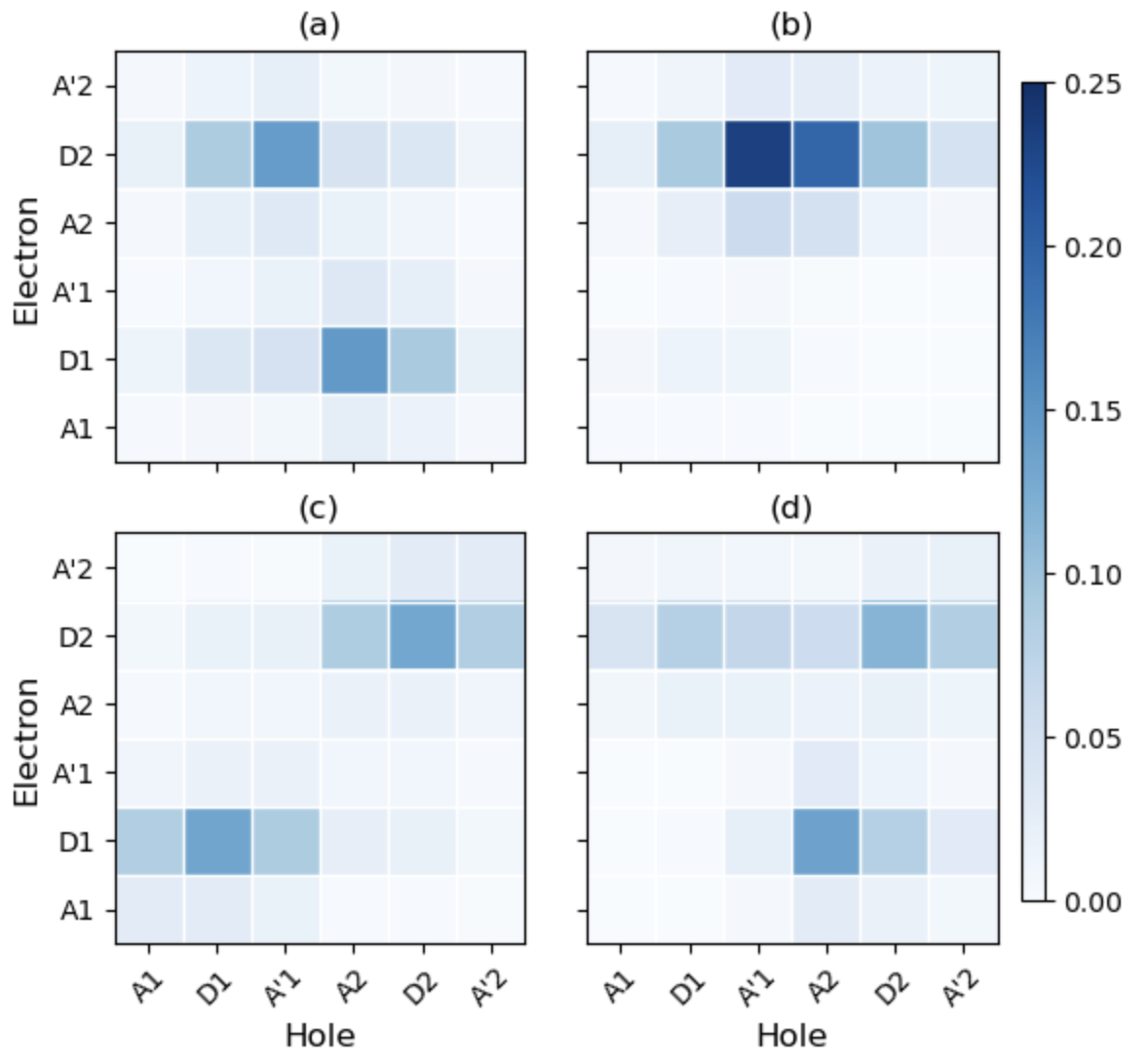}
    \caption{Electron/Hole correlation plots of the $S_1$ (top left) and $S_3$ (bottom left) states  on dimer D2, and the $S_1$ (top right) and $S_3$ (bottom right) states dimer D4 . Monomers within each dimer are fragmented into two (acceptor) end units and a core (donor) unit, denoted as $A/A'$ and $D$ respectively. Results obtained from electronic structure calculations using OT-LC-$\omega$hPBE/6-31G(d,p) and wavefunction analysis carried out using the Theodore3.2 package \cite{plasser2020theodore}. The colourbar again corresponds to the 1 particle transition density (given in terms of the electron and hole position) for the given excited state. On-diagonal elements correspond to the electron and hole being correlated on the same fragment, e.g.\ a Frenkel Exciton and off-diagonal elements correspond to Charge-Transfer excitation.   }
    \label{e/h correlation}
\end{figure}

\section{Different dimers}

\begin{figure}
    \centering
    \includegraphics[width=0.75\linewidth]{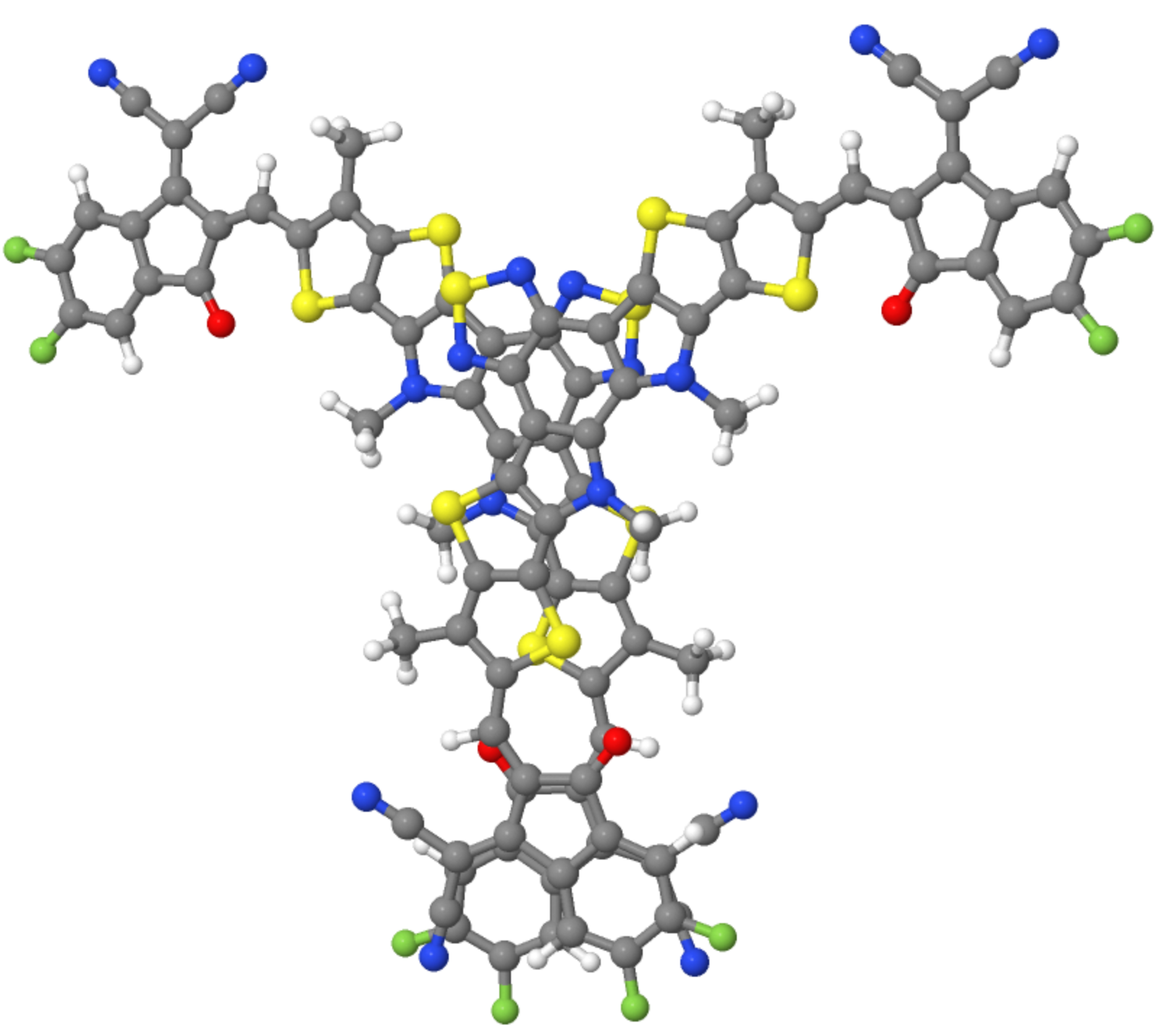}
    \caption{D1 dimer as an example of a H aggregate in Y6.}
    \label{D1_dimer}
\end{figure}

\begin{figure}
    \centering
    \includegraphics[width=0.75\linewidth]{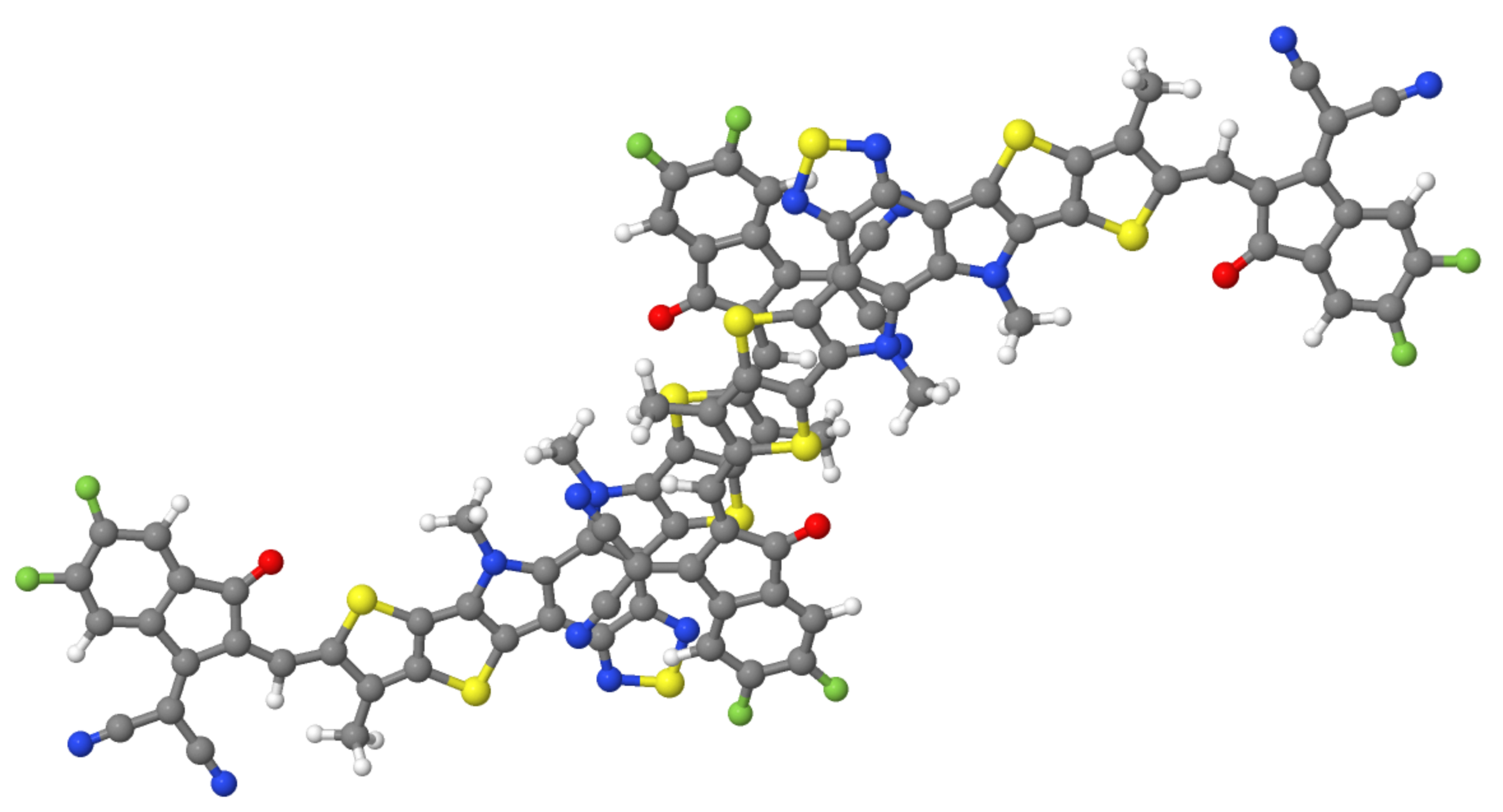}
    \caption{D2 dimer as an example of a J aggregate in Y6.}
    \label{D2_dimer}
\end{figure}

\begin{figure}
    \centering
    \includegraphics[width=0.75\linewidth]{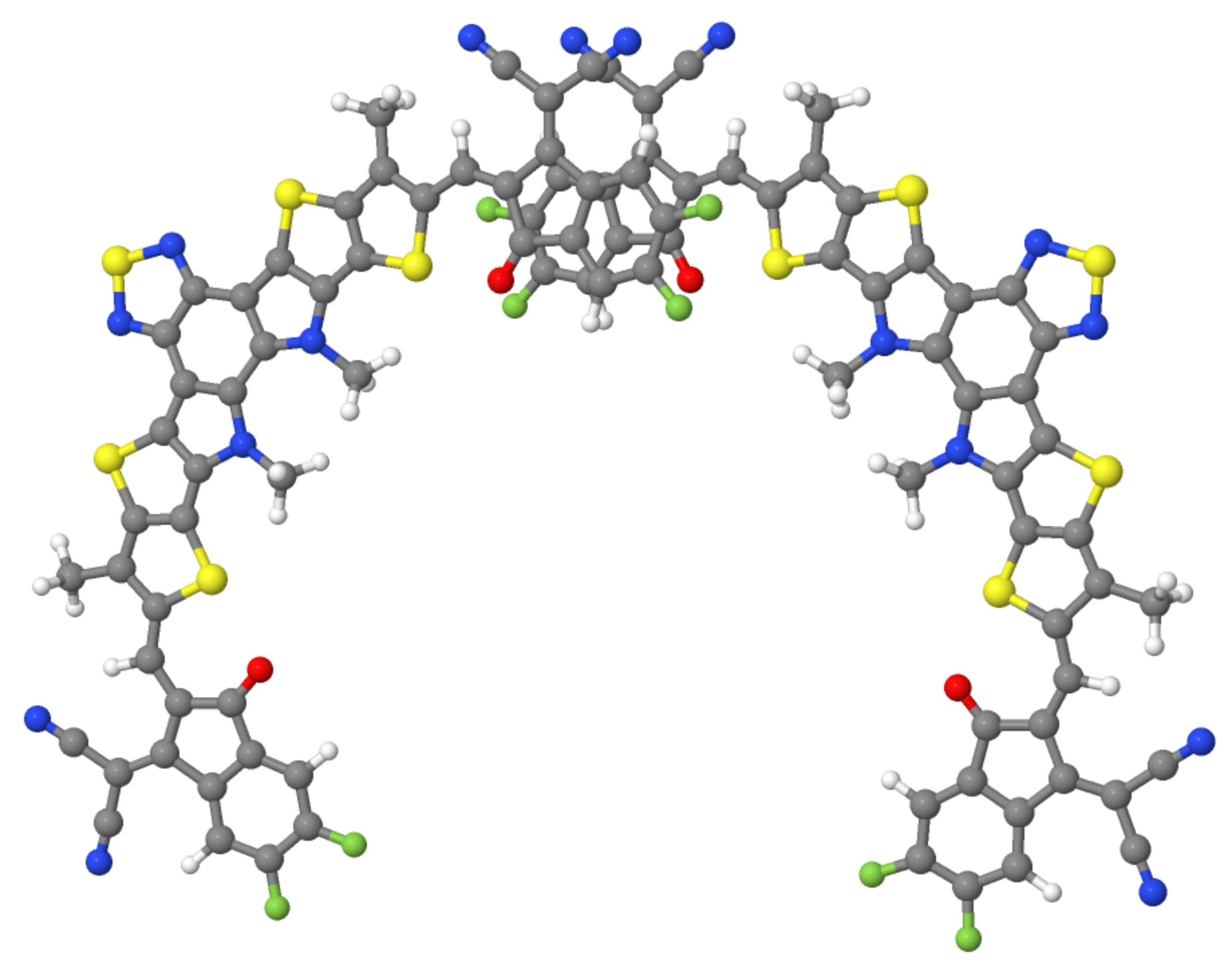}
    \caption{D4 dimer as an example of a V aggregate in Y6.}
    \label{D4_dimer}
\end{figure}

We here show the three different types of dimer found in the Y6 crystal structure namely, the H aggregate D1 dimer (Fig \ref{D1_dimer}), the J aggregate D2 dimer (Fig \ref{D2_dimer}) and the V aggregate D4 dimer (Fig \ref{D4_dimer}).

\end{document}